\documentclass[aps,prd,preprint,groupedaddress]{revtex4-1}

\usepackage{graphicx}
\usepackage{amsmath}
\usepackage{amsfonts}      % facilitate mathematical typesetting
\usepackage{hyperref}     % hyperlinks
\usepackage{multirow}
\usepackage{xcolor}
\usepackage{slashed}
\usepackage{makecell}
\allowdisplaybreaks

\newcommand{\nn}{\nonumber\\}

\begin{document}
\pagenumbering{gobble}
\title{The $H^*H^*V$  couplings from light-cone sum rules}

\author{Chao Wang}
\email[]{chaowang@nankai.edu.cn}
\affiliation{ \small Faculty of Mathematics and Physics, Huai'an University, Huaian 223001,  China}

\date{\today}

\begin{abstract}
	We present a determination of the strong charge couplings $g_{H^*H^*V}$ and magnetic couplings $f_{H^*H^*V}$ (with $H \in \{D, B\}$ and $V \in \{\rho, \omega, K^*, \phi\}$) using the framework of light-cone sum rules (LCSR). The theoretical precision is improved by establishing the leading-power hard-collinear factorization formulas with next-to-leading-order (NLO) $\alpha_s$ corrections, alongside the inclusion of power-suppressed contributions up to the next-to-next-to-leading power (NNLP) for both channels. By parameterizing the $\mathcal{O}(1/m_{H^*})$ power corrections, we extract the universal static couplings $\beta = 0.73 \pm 0.13$ and $\hat{\lambda} = 0.23 \pm 0.06 \text{ GeV}^{-1}$, highlighting their distinct heavy-quark scaling behaviors. While our investigation into SU(3) flavor symmetry breaking shows that these effects are currently overwhelmed by uncertainties in the non-perturbative vector meson distribution amplitudes, our determinations of these spin-conserving and spin-flip parameters provide non-perturbative inputs for modeling the S-wave central potentials, as well as the tensor forces and D-wave mixing effects, in exotic heavy tetraquarks. 	
\end{abstract}

\maketitle

\pagenumbering{arabic}

\section{Introduction}

A quantitative understanding of low-energy strong interactions among heavy and light hadrons remains an ongoing challenge in modern flavor physics. In the heavy quark sector, the dynamics are governed by approximate Heavy Quark Spin Symmetry (HQSS)~\cite{Wise:1992hn,Neubert:1993mb}, which naturally predicts universal behaviors for low-energy effective interactions. Within the framework of Heavy Meson Chiral Perturbation Theory (HM$\chi$PT)~\cite{Casalbuoni:1996pg,Yan:1992gz,Burdman:1992gh}, both the charge couplings $g_{H^*H^*V}$ and the magnetic couplings $f_{H^*H^*V}$ governing the interactions between two heavy vector mesons ($D^*$ or $B^*$) and a light vector meson serve as essential inputs. However, physical heavy quarks possess finite masses, and the corresponding mass corrections break this exact spin symmetry. Consequently, a comparative analysis of these couplings in both the charm and bottom sectors is highly desirable to quantitatively isolate the subleading heavy quark mass corrections and to extract the universal heavy-quark coupling parameters.

The precise evaluation of these on-shell couplings is currently of phenomenologically highly relevant. In the exploration of exclusive non-leptonic decays of heavy mesons, final state interactions (FSIs) play a significant role in generating the strong phase angles that are required to predict direct CP violation \cite{Wolfenstein:1990ks,Donoghue:1996hz, Cheng:2004ru, Cheng:2010ry}. Evaluating these intricate rescattering effects relies heavily on hadronic meson-exchange models. Moreover, these dynamic couplings serve as essential ingredients for constructing the one-boson-exchange potentials in the study of exotic hadronic molecules. The charge coupling $g_{H^*H^*V}$ dictates the central potentials in S-wave states, whereas the magnetic coupling $f_{H^*H^*V}$ governs the tensor forces and D-wave mixing effects~\cite{Tornqvist:1993ng}. Quantifying these specific coupled-channel dynamics is critical for understanding the binding mechanisms of the $D^*\bar{D}^*$ and $B^*\bar{B}^*$ molecular candidates, such as the $Z_c(4020)$ and $Z_b(10650)$ states~\cite{BESIII:2013ouc,BESIII:2013mhi,Belle:2011aa,He:2013nwa,Sun:2011uh}, as well as anticipating the HQSS partners of the doubly charmed tetraquark $T_{cc}^+$ \cite{LHCb:2021vvq, Albaladejo:2021vln, Dong:2021bvy}.

Beyond heavy quark symmetry, these couplings also provide a theoretical probe into fundamental $SU(3)$ flavor symmetry breaking \cite{Casalbuoni:1996pg}. By systematically investigating the transition channels involving strange quarks, one can quantitatively assess the symmetry departure driven by the finite strange quark mass and the asymmetric distribution amplitudes of the light vector mesons \cite{Ball:1998sk,Ball:1998ff, Ball:2007rt, Ball:2007zt, Bharucha:2015bzk, Gao:2019lta}.  A comparative analysis of these channels is highly motivated not only to estimate the central magnitude of the breaking, but more importantly, to expose the current theoretical precision limits imposed by the non-perturbative inputs.

Over the past decades, the charge coupling $g_{H^*H^*V}$ was extensively estimated using three-point QCD sum rules (QCDSR) by calculating off-shell form factors for both the $H^*$ and $V$ channels~\cite{Bracco:2007sg,Bracco:2011pg,Janbazi:2018bgv,Khosravi:2013ad,Cui:2012wk}. In Refs.~\cite{Wang:2007ci,Li:2007dv,Aliev:2021cjt}, the $g_{H^*H^*V}$ coupling was also evaluated within the framework of light-cone sum rules (LCSR)~\cite{Balitsky:1989ry,Belyaev:1994zk} at the leading order (LO) in $\alpha_s$, leaving room for further improvements through next-to-leading order (NLO) corrections. In contrast, theoretical evaluations of the magnetic coupling $f_{H^*H^*V}$ are limited in the existing literature~\cite{Li:2007dv,Cui:2012wk}. In traditional sum rule approaches, extracting this spin-flip coupling is technically challenging due to severe continuum contaminations and the complexities of isolating the associated momentum-dependent tensor structures. Complementary to the sum rule approaches, lattice QCD (LQCD) extracts the strong couplings $g_{D^*D^*\rho}$ by evaluating the hadronic matrix elements of the vector current on the discretized spacetime lattice~\cite{Can:2012tx}. Recently, LQCD simulations have advanced significantly in extracting heavy meson scattering amplitudes and coupled-channel dynamics \cite{Padmanath:2022cvl, Shrimal:2025ues,Shrimal:2026bur}. These on-shell couplings are essential inputs for such frameworks.

Building upon previous theoretical efforts, the LCSR approach provides an elegant and systematic framework to extract such hadronic coupling constants \cite{Li:2020rcg,Khodjamirian:2020mlb,Wang:2020yvi,Jin:2024zyy,Jiang:2024equ, Wang:2026sby}. To ensure high theoretical accuracy and advance our understanding of non-perturbative QCD, it is necessary to carry out the calculation beyond the leading approximation. By employing LCSR in conjunction with soft-collinear effective theory (SCET) \cite{Bauer:2000yr,Bauer:2001yt,Beneke:2002ph}, one can establish a rigorous hard-collinear factorization framework. Furthermore, the light-cone expansion allows for the systematic inclusion of power-suppressed contributions encoded in higher-twist multi-particle distribution amplitudes (DAs).

In this work, we present an updated extraction of both the charge ($g_{H^*H^*V}$) and magnetic ($f_{H^*H^*V}$) strong couplings within the LCSR framework. Specifically, we establish the hard-collinear factorization formula for the underlying correlation function at the leading power (LP) up to NLO in $\alpha_s$, detailing the computation of the one-loop hard matching coefficients. Furthermore, we incorporate the power-suppressed contributions at LO by evaluating the two- and three-particle higher-twist DAs up to the next-to-next-to-leading power (NNLP). By matching the QCD-level double spectral representations with the hadronic dispersion relations, we construct the analytical LCSR. The remainder of this paper is organized as follows. In Sec.~\ref{sec-2}, we introduce the definition of the strong couplings and derive the hadronic dispersion relation. In Sec.~\ref{sec-3}, we perform the hard-collinear factorization and derive the master formulae for the sum rules. In Sec.~\ref{sec-4}, we carry out a comprehensive numerical analysis to extract the couplings and discuss HQSS breaking effects. Sec.~\ref{sec-5} is reserved for a summary of our main observations. Finally, the explicit expressions for the vector QCD DAs and one-loop Feynman diagrams are collected in Appendix~\ref{appendix-A} and \ref{appendix-B}, respectively.

\section{Theory summary for the $H^*H^*V$ couplings}
\label{sec-2}

The effective Lagrangian describing the strong interactions between heavy vector mesons and light vector mesons within the framework of HM$\chi$PT can be written as~\cite{Cheng:2004ru}
\begin{align}
	\mathcal{L}=
	-ig_{H^*H^*V} H_i^{*\nu\dagger} (\overrightarrow{\partial}_\mu -\overleftarrow{\partial}_\mu)  H_\nu^{*j} (V^\mu)^i_j-4if_{H^*H^*V} H_{i\mu}^{*\dagger}   (\overrightarrow{\partial}^\mu V^\nu-\overrightarrow{\partial}^\nu V^\mu)^i_j H_\nu^{*j}\,,
\end{align}
where $H_\mu^*$ denotes the heavy vector meson triplet, explicitly $H^*=(D^{*0}, \, D^{*+},\, D_s^{*+})$ for the charm sector and $H^*=(B^{*-}, \, \bar{B}^{*0},\, \bar{B}_s^{*0})$ for the bottom sector. The light vector mesons are parameterized by the $3\times 3$ matrix $V_\mu$,
\begin{align}
	V= 
	\begin{pmatrix}
		(\rho^0+\omega)/\sqrt{2} & \rho^+ & K^{*+}\\
		\rho^- & -(\rho^0-\omega)/\sqrt{2} & K^{*0}\\
		K^{*-} & \overline{K}^{*0} &\phi\\
	\end{pmatrix}\,.
\end{align}

From this Lagrangian, the corresponding hadronic matrix element is parameterized as
\begin{align}
	&\langle D^{*0}(q,\eta_2) \rho^+(p,\epsilon) |\mathcal{L} | D^{*+}(p+q,\eta_1) \rangle \nn
	&= -2g_{D^{*+}D^{*0}\rho^+} \,  (q\cdot \epsilon^*)\,(\eta_1 \cdot \eta_2^* ) -4f_{D^{*+}D^{*0}\rho^+} \big[(p\cdot \eta_1) \, (\epsilon^*  \cdot \eta_2^*)-(p \cdot \eta_2^*) \, (\epsilon^* \cdot \eta_1)\big]\,.
	\label{kinematic}
\end{align}
The charge coupling $g_{H^*H^*V}$ stems from the spin-conserving (charge-like) interactions of the light degrees of freedom within the heavy meson, whereas the magnetic coupling $f_{H^*H^*V}$ originates from the spin-flip (magnetic dipole-like) interactions. As explicitly shown in the matrix element, the kinematic prefactor associated with $f_{H^*H^*V}$ is proportional to the light meson momentum $p \sim \mathcal{O}(\Lambda_{\text{QCD}})$, rendering its kinematic contribution suppressed by a factor of $\mathcal{O}(\lambda)$ (where $\lambda = \Lambda_{\text{QCD}}/m_Q$) relative to the charge coupling term (which couples to the heavy meson momentum $q \sim \mathcal{O}(m_Q)$). However, to yield an unsuppressed physical scattering amplitude as dictated by HQSS, the coupling parameter $f_{H^*H^*V}$ must be power-enhanced to offset this kinematic suppression (as indicated by Eq. (\ref{eq:HQSS_breaking})).

In the strict heavy quark and chiral limit, HQSS dictates that the charge and magnetic couplings are related to the universal static couplings $\beta$ and $\hat{\lambda}$ via $g_{H^*H^*V} = g_V \beta /\sqrt{2}$ and $f_{H^*H^*V} = g_V \hat{\lambda} m_{H^*}/\sqrt{2} $, respectively, with $g_V = m_\rho/f_\pi$. To account for the heavy quark symmetry breaking effects induced by the finite heavy quark masses, we parameterize the physical couplings with the first-order $1/m_{H^*}$ power corrections~\cite{Belyaev:1994zk},
\begin{align}
	g_{H^*H^*V} = \frac{g_V}{\sqrt{2}} \,\beta \Big(1 + \frac{\delta_a}{m_{H^*}}\Big) \,, \quad f_{H^*H^*V} = \frac{ g_V}{\sqrt{2}}\, \hat{\lambda}\, m_{H^*} \Big(1 + \frac{\delta_b}{m_{H^*}}\Big)\,,
	\label{eq:HQSS_breaking}
\end{align}
where the parameter $\delta_{a(b)}$ quantifies the mass-dependent symmetry breaking effect.

For the specific couplings, we use the simplified notation:
\begin{align}
	g[f]_{D^{*}D^{*}\rho}&=g[f]_{D^{*+}D^{*0}\rho^+} \,, \quad g[f]_{D^{*}D^{*}\omega}=g[f]_{D^{*+}D^{*+}\omega}\,, \nn
	g[f]_{D_s^{*}D^{*}K^*}&=g[f]_{D_s^{*+}D^{*0}K^{*+}} \,, \quad
	g[f]_{D_s^{*}D_s^{*}\phi}=g[f]_{D_s^{*+}D_s^{*+}\phi}  \,.
\end{align}
The couplings for the other charge states are related to these by isospin symmetry. For the $B^*B^*V$ couplings, the corresponding relations are obtained by the replacement $D^*\to \bar{B}^*$. Considering ideal $SU(3)$ flavor symmetry, these couplings obey the strict algebraic proportionality:
\begin{align}
	g[f]_{D^{*}D^{*}\rho}=\sqrt{2}g[f]_{D^{*}D^{*}\omega} = g[f]_{D_s^{*}D^{*}K^*}=g[f]_{D_s^{*}D_s^{*}\phi}\,. 
\end{align}

To extract these couplings via LCSR, we construct the vacuum-to-vector-meson correlation functions utilizing local interpolating currents for the heavy vector mesons,
\begin{align}
	\Pi(p,q) &= i\int d^4x\,e^{iq\cdot x} \langle V(p,\epsilon) |T\{\bar{q}_1(x) \gamma_{\mu\perp} Q(x), \, \bar{Q}(0) \gamma^\mu_\perp q_2(0) \} | 0\rangle \nn
	&= (n\cdot q\,\bar{n}\cdot \epsilon^*)\, \Pi_g(p,q) + \cdots\,, \nn
	\Pi_{\mu\perp}(p,q) &= i\int d^4x\,e^{iq\cdot x} \langle V(p,\epsilon) |T\{\bar{q}_1(x) \slashed{\bar{n}} Q(x), \, \bar{Q}(0) \gamma_{\mu\perp}  q_2(0) \} | 0\rangle \nn
	&= (\bar{n}\cdot p  \,\epsilon_{\mu\perp}^*)\, \Pi_f(p,q) + \cdots\,, \label{cfun}
\end{align}
where the ellipses represent other independent Lorentz structures that are irrelevant to our analysis. Here  $Q \in \{c, b\}$ represents the heavy quark field, and $q_{1,2} \in \{u, d, s\}$ denote the corresponding light quark fields. The transverse gamma matrix is given by
\begin{align}
	\gamma_{\mu\perp}=\gamma_\mu-\frac{\slashed{\bar{n}}}{2} n_\mu- \frac{\slashed{n}}{2} \bar{n}_\mu\,,
\end{align}
with the light-cone vectors satisfying $n\cdot \bar{n}=2$ and $n^2=\bar{n}^2=0$. 
Taking advantage of the standard definition for the heavy vector meson decay constant,
\begin{align}
	\langle 0| \bar{q} \gamma_{\mu} Q| H^*(q,\eta) \rangle =f_{H^*} m_{H^*} \eta_{\mu} \,,
\end{align}
and inserting a complete set of intermediate hadronic states with the $H^*$ quantum numbers into the correlation functions, we can isolate the targeted scalar invariant amplitudes to obtain the following hadronic double dispersion relations:
\begin{align}
	\Pi_g(p,q) &= -\frac{2g_{H^*H^*V} f_{H_1^*} f_{H_2^*} m_{H_1^*} m_{H_2^*} }{[m_{H_1^*}^2-(p+q)^2] [m_{H_2^*}^2-q^2]} + \iint_\Sigma ds_1ds_2 \frac{\rho_g^h(s_1,s_2)}{(s_1-(p+q)^2)(s_2-q^2)} +\cdots\,, \nn
	\Pi_f(p,q) &= \frac{4f_{H^*H^*V} f_{H_1^*} f_{H_2^*} m_{H_1^*} m_{H_2^*} }{[m_{H_1^*}^2-(p+q)^2] [m_{H_2^*}^2-q^2]} +\iint_\Sigma ds_1ds_2 \frac{\rho_f^h(s_1,s_2)}{(s_1-(p+q)^2)(s_2-q^2)} +\cdots\,, \label{hadronic}
\end{align}
where $\Sigma$ stands for the parton-hadron duality region. The function $\rho_{g,f}^h(s_1,s_2)$ denotes the spectral density of higher resonances and continuum states, and the ellipsis represents the subtraction polynomials, which identically vanish upon applying the double Borel transformation.

\section{LCSR for the $H^*H^*V$ couplings}
\label{sec-3}

\subsection{The hard-collinear factorization at leading power}

To compute the invariant amplitudes at the partonic level, we match the underlying full QCD correlation functions onto the effective operators in SCET. We treat the propagating heavy quark as highly virtual and consistently adopt the $\overline{\text{MS}}$ scheme for the heavy quark mass $m_Q$ throughout the calculation. Furthermore, we employ the following power counting scheme in the operator product expansion (OPE) region:
\begin{align}
	&|(p+q)^2-m_Q^2| \sim |q^2-m_Q^2| \sim m_Q^2\,, \quad  \bar{n}\cdot p\sim m_Q, \quad n\cdot p\sim \lambda^2 \,m_Q\,,\nn
	&   \epsilon \sim (\bar{n}\cdot \epsilon, \, n\cdot \epsilon, \, \epsilon_\perp) \sim(\lambda^{-1}, \, \lambda, \, 1), \quad m_V\sim f_V^\parallel\sim f_V^\perp \sim \lambda\, m_Q\,.
\end{align}

At the leading power, matching the full QCD amplitudes onto the SCET operators yields the factorized forms,
\begin{align}
	\Pi(p,q) &= \int_0^1 du' \, \mathbb{C}_g(p,q,u') \langle V(p, \epsilon) | O_V(u') | 0 \rangle^{(0)}\,, \nn
	\Pi_{\mu\perp}(p,q) &= \int_0^1 du' \, \mathbb{C}_f(p,q,u') \langle V(p, \epsilon) | O_{T,\mu\perp}(u') | 0 \rangle^{(0)}\,,
\end{align}
where $\mathbb{C}_{g,f}$ denote the short-distance Wilson coefficients, and the bare SCET operators are defined as
\begin{align}
	O_V(u') &= \frac{\bar{n}\cdot p}{2\pi}\int d\tau \,e^{-iu' \tau \bar{n}\cdot p} \bar{\chi}(\tau \bar{n}) \slashed{\bar{n}} \chi(0) \,, \nn
	O_{T,\mu\perp}(u') &= \frac{\bar{n}\cdot p}{2\pi}\int d\tau \,e^{-iu' \tau \bar{n}\cdot p} \bar{\chi}(\tau \bar{n}) \slashed{\bar{n}} \gamma_{\mu\perp} \chi(0) \,.
\end{align}

By employing the standard definitions of the leading-twist vector meson DAs in SCET,
\begin{align}
	\langle V(p,\epsilon)|\bar{\chi}_1(x_+) \slashed{\bar{n}} \chi_2(0)|0\rangle &=f_V^\parallel m_V \,\bar{n}\cdot \epsilon^* \int_0^1du \, e^{iup\cdot x_+} \,\phi_2^{\parallel}(u)\,, \nn
	\langle V(p,\epsilon)|\bar{\chi}_1(x_+) \slashed{\bar{n}} \gamma_{\mu\perp} \chi_2(0)|0\rangle &=f_V^\perp  \,\bar{n}\cdot p\,\epsilon^*_{\mu\perp} \int_0^1du \, e^{iup\cdot x_+} \,\phi_2^{\perp}(u)\,,
\end{align}
with $x_+=n\cdot x\, \bar{n}/2$, the convolution integrals over the momentum fraction $u'$ explicitly yield the $\delta(u-u')$ distribution. We can then project out the designated Lorentz structures $(n\cdot q\,\bar{n}\cdot \epsilon^*)$ and $(\bar{n}\cdot p \,\epsilon_{\mu\perp}^*)$ defined in Eq.~(\ref{cfun}) to isolate the LO scalar invariant amplitudes
\begin{align}
	\Pi_g^{(0)}(p,q) &= f_V^\parallel m_V \int_0^1du \, \phi_2^{\parallel}(u) H_g^{(0)}\big((p+q)^2,q^2,u\big)\,, \nn
	\Pi_f^{(0)}(p,q) &= f_V^\perp \int_0^1du \, \phi_2^{\perp}(u) H_f^{(0)}\big((p+q)^2,q^2,u\big)\,.
\end{align}
Here, the purely scalar LO hard kernels are given by
\begin{align}
	H_g^{(0)}(u) = \frac{1}{u(p+q)^2+\bar{u}q^2-m_Q^2} \,, \quad H_f^{(0)}(u) = \frac{-m_Q}{u(p+q)^2+\bar{u}q^2-m_Q^2} \,.
\end{align}
At the partonic level, the overall mass factor $m_Q$ appearing in the numerator of the magnetic hard kernel $H_f^{(0)}$ originates from the leading-power expansion of the highly virtual heavy-quark propagator. This $m_Q$ factor acts as the dynamical origin that counterbalances the $\mathcal{O}(\lambda)$ kinematic suppression of the spin-flip interaction, fulfilling the scaling compensation anticipated in Eq.~(\ref{eq:HQSS_breaking}).

\begin{figure}[htpb]
	\begin{center}
		\includegraphics[width=0.8 \columnwidth]{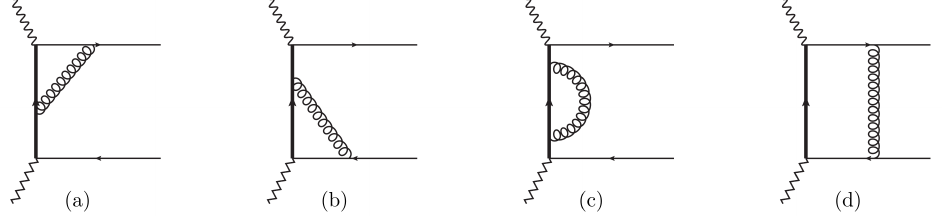} \\
		\caption{One-loop Feynman diagrams for the QCD matrix element $F(p,q,u)$ and $F_{\mu\perp}(p,q,u)$.}
		\label{twist-2-NLO}
	\end{center}
\end{figure}

To evaluate the $\mathcal{O}(\alpha_s)$ radiative corrections, we calculate the one-loop Feynman diagrams in full QCD as illustrated in Fig.~\ref{twist-2-NLO} for both correlation functions. Specifically, Fig.~\ref{twist-2-NLO}(a) and \ref{twist-2-NLO}(b) represent the vertex corrections, Fig.~\ref{twist-2-NLO}(c) denotes the quark self-energy correction, and Fig.~\ref{twist-2-NLO}(d) shows the one-loop box diagram contribution. The explicit results are listed in  Appendix~\ref{appendix-B}.
Since dimensional regularization is employed to regulate both the  UV and  IR divergences, the bare one-loop integrals of the SCET operators are purely scaleless and vanish identically. Consequently, the IR divergences of the full QCD amplitude are canceled by the UV renormalization factor of the SCET operator. After this subtraction,  the one-loop hard matching coefficients are as follows:
\begin{align}
	H_g^{(1)}
	&=\bigg\{2\bigg[ \frac{(r_1+1)(r_2+r_3)-r_1^2-2r_2r_3-1}{(r_1-r_2)(r_1-r_3)} L(r_1) 
	+\frac{1-r_1+r_2-r_3}{(r_1-r_2)(r_2-r_3)}(1-r_2) L(r_2)\nn 
	&\quad+\frac{1-r_1+r_3-r_2}{(r_1-r_3)(r_3-r_2)} (1-r_3)L(r_3) -\frac{3}{1-r_1} -\frac{3}{2} \bigg]  \,L  \nn 
	&\quad+\frac{2(1+r_1^2-r_3-r_1(r_2+r_3)+r_2(2r_3-1))}{(r_1-r_2)(r_1-r_3)}  L_2(r_1) \nn
	&\quad-\frac{2(1-r_2)(1-r_1+r_2-r_3)}{(r_1-r_2)(r_2-r_3)}  L_2(r_2)
	-\frac{2(1-r_3)(1-r_1-r_2+r_3)}{(r_1-r_3)(r_3-r_2)} L_2(r_3) \nn
	&\quad+\frac{5r_1^2+r_2r_3+2r_1(1-r_2-r_3)}{r_1^2(r_1-r_2)(r_1-r_3)} (1-r_1)^2L(r_1) +\frac{1-9r_1+6r_1^2}{r_1(1-r_1)}\nn
	&\quad+\frac{-2(1-r_1)+r_3(1-r_2)+r_2(2r_1+r_2-3)}{(r_1-r_2)r_2(r_2-r_3)} (1-r_2)L(r_2) \nn
	&\quad+\frac{-2(1-r_1)+r_2(1-r_3)+r_3(2r_1+r_3-3)}{(r_1-r_3)r_2(r_3-r_2)} (1-r_3)L(r_3) \bigg\} H_g^{(0)} \,, \label{hard-fun-g} \\
	H_f^{(1)}
	&= \bigg\{(-2)\bigg[ \frac{1-r_2}{r_1-r_2} \big(L(r_1) -L(r_2)\big) +\frac{1-r_3}{r_1-r_3}\big(L(r_1) -L(r_3)\big) +\frac{3}{1-r_1}  \bigg]  L \nn 
	&+2\Big(\frac{1-r_2}{r_1-r_2}+\frac{1-r_3}{r_1-r_3} \Big)  L_2(r_1) 
	-\frac{2(1-r_2)}{r_1-r_2}   L_2(r_2)
	-\frac{2(1-r_3)}{r_1-r_3}   L_2(r_3) \nn
	&+\frac{-r_2+4r_1r_2(1-r_3) +r_3-r_1^2(4-r_2-3r_3)}{r_1(r_1-r_2)(r_1-r_3)} L(r_1) 
	+\frac{1-r_2^2}{r_2(r_1-r_2)} L(r_2) \nn &-\frac{(1-3r_3)(1-r_3)}{r_3(r_1-r_3)} L(r_3) -\frac{8}{1-r_1} -3\bigg\} H_f^{(0)} \,,
	\label{hard-fun-f}
\end{align}
where the dimensionless kinematic variables are defined as $r_1=(up+q)^2/m_Q^2$, $r_2=q^2/m_Q^2$, and $r_3=(p+q)^2/m_Q^2$. We have also introduced the auxiliary functions
\begin{align}
	L=\ln\frac{\mu^2}{m_Q^2}\,, \quad L(r_i)=\ln(1-r_i)\,, \quad L_2(r_i)=\ln^2(1-r_i)+{\rm Li}_2(r_i)\,.
\end{align} 
The final OPE results for the invariant amplitudes at LP up to $\mathcal{O}(\alpha_s)$ can be cast into a unified factorization formula
\begin{align}
	\Pi_I^{\rm LP}(p,q) = \mathcal{C}_I \int_{0}^{1}du \,\phi_{2,I}(u) \Big[H_I^{(0)}\big((p+q)^2,q^2,u\big) +\frac{\alpha_sC_F}{4\pi} H_I^{(1)}\big((p+q)^2,q^2,u\big) \Big] \,,
	\label{ope-nlo}
\end{align}
where the label $I \in \{g, f\}$ denotes the channels for extracting the charge and magnetic couplings, respectively. The overall normalization coefficients are given by $\mathcal{C}_g = f_V^\parallel m_V$ and $\mathcal{C}_f = f_V^\perp$, alongside the specific leading-twist distribution amplitudes $\phi_{2,g} = \phi_2^\parallel$ and $\phi_{2,f} = \phi_2^\perp$, and the one-loop kernels $H_I^{(1)}$ given in Eqs.~(\ref{hard-fun-g}) and (\ref{hard-fun-f}).

The renormalization scale independence of the resulting $\mathcal{O}(\alpha_s)$ factorization formulae can be verified by utilizing the standard ERBL evolution equations~\cite{Efremov:1979qk,Lepage:1980fj} for the leading-twist longitudinal and transverse DAs,
\begin{align}
	&\quad \frac{d}{d\ln\mu^2} \phi_2^\parallel(u,\mu) \nn&= \frac{\alpha_sC_F}{2\pi}  \int_0^1 du'  \phi_2^\parallel(u',\mu) \, \Big[\frac{\bar{u}}{\bar{u}'} \Big(1+\frac{1}{u-u'} \Big)\theta(u-u') +\frac{u}{u'} \Big(1+\frac{1}{u'-u}\Big)\theta(u'-u)\Big]_+ \,,\nn
	&\quad\frac{d}{d\ln\mu^2} \big[f_V^\perp(\mu) \phi_2^\perp(u,\mu)\big] \nn&=\frac{\alpha_sC_F}{2\pi}  \int_0^1du'\, \big[f_V^\perp(\mu) \phi_2^\perp(u',\mu)\big] \biggl\{\Big[\frac{\bar{u}}{\bar{u}'} \frac{\theta(u-u')}{u-u'} +\frac{u}{u'} \frac{\theta(u'-u)}{u'-u} \Big]_+  -\frac{1}{2}\delta(u-u') \biggr\}\,.
\end{align}
For both the charge and magnetic channels, the renormalization scale dependence of the respective DAs cancels against the combined $\mu$-dependence of the NLO matching kernels $H_I^{(1)}$ and the running heavy quark mass $m_Q(\mu)$ inherently present in the LO coefficients $H_I^{(0)}$.

\subsection{Power-suppressed contributions at NLP and NNLP}

To incorporate the power-suppressed higher-twist corrections, we employ the light-cone expansion of the heavy-quark propagator in the background gluon field~\cite{Balitsky:1987bk}. Inserting this expansion into the correlation functions and utilizing the vector meson DAs up to twist-4, we can isolate the invariant amplitudes beyond the leading power. To strictly preserve the mathematical consistency of the light-cone expansion and avoid the spurious kinematic singularities encountered in previous literature, we employ the asymptotic forms of the higher-twist DAs exclusively for the singular terms. A detailed discussion on safely eliminating the non-vanishing boundary terms during integration by parts is relegated to Appendix~\ref{appendix-A}.

Following this procedure, we obtain the power-suppressed invariant amplitudes at NLP (governed by twist-3 DAs) and NNLP (incorporating both twist-4 DAs and kinematic $\mathcal{O}(m_V^2)$ corrections). For the charge channel, the results are identically free from genuine three-particle twist-3 contributions due to Lorentz projection, 
\begin{align}
	\Pi_g^{(0),\rm NLP}(p,q) &= -f_V^\perp m_V^2 m_Q  \int_0^1du\, \frac{\psi_3^\parallel(u)}{\Delta^2} \,,\nn
	\Pi_g^{(0),\rm NNLP}(p,q) &= f_V^\parallel m_V^3 \int_0^1du\,
	\bigg[\frac{u\bar{u}}{\Delta^2} \phi_2^{\parallel}(u)  +  \frac{1}{4}\Big(\frac{1}{\Delta^2}- \frac{2m_Q^2}{\Delta^3}\Big) \phi_4^{\parallel}(u)  
	+ 2 \Big(\frac{1}{\Delta^2} - \frac{2 m_Q^2}{\Delta^3} \Big) \psi_4^{\parallel,(2)}(u)\bigg]\,.
	\label{ope-nlp-g}
\end{align}
where the denominator is defined as $\Delta = u(p+q)^2+\bar{u}q^2-m_Q^2$.
Conversely, for the magnetic channel, the genuine multi-parton effects inherently survive the Lorentz projection at NLP. The corresponding amplitudes read
\begin{align}
	\Pi_f^{(0),\rm NLP}(p,q) &= -f_V^\parallel m_V  \int_0^1du\, \bigg[\frac{u}{\Delta} \phi_3^\perp(u) -\frac{m_Q^2}{4\Delta^2}  \psi_3^\perp(u) +\frac{1}{\Delta}  \bar{\Phi}'_3(u) \bigg] \,,\nn
	\Pi_f^{(0),\rm NNLP}(p,q) &= -f_V^\perp m_Q  m_V^2 \int_0^1 du  \bigg[ \frac{u\bar{u}}{\Delta^2} \phi_2^\perp(u) - \frac{ m_Q^2 }{2\Delta^3} \phi_4^\perp(u) + \frac{u}{\Delta^2} \psi_4^{\perp,(1)}(u) \bigg] \,,
	\label{ope-nlp-f}
\end{align}
where the function $\bar{\Phi}'_3(u) \equiv d\bar{\Phi}_3(u)/du$ encapsulates the three-particle twist-3 distribution amplitude, with 
\begin{align}
	\bar{\Phi}_3(u) =  \int_0^u \! d\alpha_q \int_{u-\alpha_q}^{1-\alpha_q} \! d\alpha_g \, \frac{u-\alpha_q}{\alpha_g^2} \, \big[ \Phi^\parallel_3 + \tilde{\Phi}^\parallel_3 \big](\alpha_q, \, 1-\alpha_q-\alpha_g, \, \alpha_g) \,.
\end{align}

\subsection{Double spectral density and the light-cone sum rules}

The complete OPE result can be recast into the form of a double dispersion relation,
\begin{align}
	\Pi_I^{\rm OPE}(p,q)&= \!\iint \frac{ds_1ds_2 }{[s_1-(p+q)^2](s_2-q^2)}
	\Big[\rho_I^{(0),\rm LP}(s_1,s_2) + \rho_I^{(0),\rm NLP}(s_1,s_2)\nn
	&\hspace{4cm} +	\rho_I^{(0),\rm NNLP}(s_1,s_2) +\frac{\alpha_sC_F}{4\pi}\rho_I^{(1),\rm LP}(s_1,s_2)\Big]\,,
	\label{spectral-rep}
\end{align}
where the double spectral densities are obtained by taking the imaginary parts, $\rho(s_1,s_2) = \frac{1}{\pi^2} {\rm Im}_{s_2} {\rm Im}_{s_1} \Pi(s_1,s_2)$. To extract the leading-order density $\rho^{(0)}$, we perform a Taylor expansion of the DAs $\phi(u)=\sum c_ku^k$, which maps the multi-pole structures into derivatives of Dirac delta functions. The general term in $\Pi^{(0)}$ is $\int_0^1du \frac{\phi(u)}{\Delta^n}$, yielding the following double spectral density{\tiny }
\begin{align}
	\rho_n^{(\phi)}&=\frac{1}{\pi^2} {\rm Im}_{s_2}  {\rm Im}_{s_1}\int_0^1du \,\frac{\phi(u)}{(us_1+\bar{u}s_2-m_Q^2)^n} \nn
	&=\sum_{k=0} \frac{(-1)^{k+1}}{\Gamma(n)\Gamma(k+1)} \frac{d^{n-1}}{(dm_Q^2)^{n-1}}c_k(s_1-m_Q^2)^k \delta^{(k)}(s_1-s_2)\theta(s_1-m_Q^2)\,.
	\label{nlp-density}
\end{align}
Accordingly, the various components are given by
\begin{align}
	\rho_g^{(0),\rm LP}(s_1,s_2)&=f_V^\parallel m_V\,\rho_1^{(\phi_2^\parallel)} \,, \quad \rho_g^{(0),\rm NLP}(s_1,s_2)=-f_V^\perp m_V^2 m_Q\,\rho_2^{(\psi_3^\parallel)} \,, \nn
	\rho_g^{(0),\rm NNLP}(s_1,s_2)&=f_V^\parallel m_V^3\, \Big[ \rho_2^{(u\bar{u}\phi_2^\parallel)} +\frac{1}{4} \rho_2^{(\phi_4^\parallel)} +2\rho_2^{(\psi_4^{\parallel,(2)})} -\frac{m_Q^2}{2} \rho_3^{(\phi_4^\parallel)} -4m_Q^2 \rho_3^{(\psi_4^{\parallel,(2)})}\Big]\,,
\end{align}
for charge channel and
\begin{align}
	\rho_f^{(0),\rm LP}(s_1,s_2)&=-f_V^\perp m_Q\,\rho_1^{(\phi_2^\perp)} \,,\quad   \rho_f^{(0),\rm NLP}(s_1,s_2)=-f_V^\parallel m_V\,  \Big[\rho_1^{(u\phi_3^\perp)}-\frac{m_Q^2}{4} \rho_2^{(\psi_3^\perp)} + \rho_1^{(\bar{\Phi}'_3)}\Big] \,, \nn
	\rho_f^{(0),\rm NNLP}(s_1,s_2)&=-f_V^\perp m_Q m_V^2\, \Big[ \rho_2^{(u\bar{u}\phi_2^\perp)} -\frac{m_Q^2}{2} \rho_3^{(\phi_4^\perp)} + \rho_2^{(u\psi_4^{\perp,(1)})}\Big]\,,
\end{align}
for magnetic channel.

Following the procedure established in Refs.~\cite{Li:2020rcg,Khodjamirian:2020mlb,Wang:2026sby,Khodjamirian:1999hb}, we derive the NLO spectral density. In this evaluation, employing the asymptotic form of the twist-two DA yields a compact and manageable analytical expression, as the coupled corrections of $\mathcal{O}(\alpha_s a_n^\parallel)$ are numerically marginal. To facilitate the phase space integration and transparently isolate the diagonal singularities, we introduce the dimensionless variables
\begin{align}
	r=\frac{s_2-m_Q^2}{s_1-m_Q^2}\,, \quad \sigma=\frac{s_1}{m_Q^2}+\frac{s_2}{m_Q^2}-2\,.
\end{align}
The NLO spectral densities can thus be written as
\begin{align}
	 \rho_g^{(1),\rm LP}(s_1,s_2)
	&= 3\frac{f_V^\parallel m_V}{m_Q^2}\frac{(r+1)^2}{\sigma} \biggl\{ (3L+1) \delta(\sigma) \delta(r-1) +\theta(\sigma)\delta(r-1) \Big[ -2{\rm Li}_2\Big(-\frac{\sigma}{2} \Big) \nn
	&\quad - \ln\frac{\sigma+2}{2} \ln\frac{\sigma}{2} 
	+\frac{24+32\sigma+10\sigma^2+\sigma^3}{2(\sigma+2)^3} \ln\frac{\sigma}{2} -\frac{1}{2}\ln\frac{\sigma+2}{2} -\frac{3}{(\sigma+2)^2} \nn 
	&\quad -\frac{2\pi^2}{3}\Big] +\theta(\sigma) \theta(r) \frac{r}{r+1} \Big(-4\ln r-2\ln\frac{\sigma+r+1}{r\sigma+r+1} + \frac{r+1}{\sigma+r+1} \nn 
	&\quad-\frac{r+1}{r\sigma+r+1}\Big) \frac{d^3}{dr^3} \ln|r-1|\biggr\}\,, \\
	\rho_f^{(1),\rm LP}(s_1,s_2)
	&=-3\frac{f_V^\perp}{m_Q}\frac{(r+1)^2}{\sigma} \biggl\{ (3L+4) \delta(\sigma) \delta(r-1) +\theta(\sigma)\delta(r-1) \Big[ -2{\rm Li}_2\Big(-\frac{\sigma}{2} \Big) \nn
	& \quad - \ln\frac{\sigma+2}{2} \ln\frac{\sigma}{2} 
	+\frac{(\sigma+4)(3\sigma+4)}{2(\sigma+2)^2} \ln\frac{\sigma}{2} -\frac{1}{2}\ln\frac{\sigma+2}{2} -\frac{1}{\sigma+2} -\frac{2\pi^2}{3}\nn &\quad -\frac{3}{4}(L+1)\Big]  +\theta(\sigma) \theta(r) \frac{r}{r+1} \Big(-4\ln r-2\ln\frac{\sigma+r+1}{r\sigma+r+1} + \frac{r+1}{\sigma+r+1} \nn&\quad +\frac{r+1}{r\sigma+r+1}-2\Big) \frac{d^3}{dr^3} \ln|r-1|\biggr\}\,.
\end{align}

By equating the OPE representations in Eq.~(\ref{spectral-rep}) with the hadronic dispersion relations in Eq.~(\ref{hadronic}), we invoke the quark-hadron duality ansatz to subtract the continuum contributions. Subsequently, performing a double Borel transformation with respect to the momenta $(p+q)^2\to M_1^2$ and $q^2\to M_2^2$ (setting $M_1^2=M_2^2=2M^2$), which exponentially suppresses the excited states and continuum contributions, we arrive at the unified LCSR for both the charge and magnetic couplings,
\begin{align}
	C_I f_{H_1^*} f_{H_2^*} = \frac{\mathcal{N}_I}{m_{H_1^*} m_{H_2^*}} \, e^{\frac{m_{H_1^*}^2+m_{H_2^*}^2}{2M^2}} \Big[\mathcal{F}_I^{(0), \rm LP} +\mathcal{F}_I^{(0), \rm NLP} +\mathcal{F}_I^{(0), \rm NNLP} +\frac{\alpha_s C_F}{4\pi} \mathcal{F}_I^{(1),\rm LP}\Big] \,,
	\label{result}
\end{align}
with the respective couplings denoted as $C_g = g_{H^*H^*V}$ and $C_f = f_{H^*H^*V}$, alongside the  numerical factors $\mathcal{N}_g = -1/2$ and $\mathcal{N}_f = 1/4$.

Unlike the leading-order contributions, which are strictly localized on the diagonal of the $(s_1, s_2)$ plane and thus depend exclusively on the diagonal intercept $s_0$, the explicit analytical expression for the NLO invariant function $\mathcal{F}_I^{(1),\rm LP}$ depends on the two-dimensional geometric shape of the chosen duality region $\Sigma$.  The general parameterization of the boundary $\Sigma$ is given by~\cite{Balitsky:1989ry}
\begin{align}
	\Big(\frac{s_1}{s_*}\Big)^\alpha +\Big(\frac{s_2}{s_*}\Big)^\alpha \leq 1\,,
\end{align}
where $s_*$ is adjusted to provide equal diagonal intervals $s_1=s_2 \leq s_0$ for different values of $\alpha$. In the following, we choose the triangle region ($\alpha=1$, $s_*=2s_0$) as our default choice, which allows us to obtain the analytical results of the $r$-integral directly. We will further discuss the errors associated with the dual region selection using different boundary parameters (concave with $\alpha=1/2$, $s_*=4s_0$ and convex with $\alpha=2$, $s_*=\sqrt{2}s_0$) in the subsequent numerical analysis.  The analytical expressions of the invariant functions $\mathcal{F}_g$  for the charge channel are summarized as follows:
\begin{align}
	\mathcal{F}_g^{(0), \rm LP} &= -f_V^\parallel m_V M^2\big(e^{-\frac{m_Q^2}{M^2}} -e^{-\frac{s_0}{M^2}} \big) \phi_2^\parallel(1/2)\,, \nn
	\mathcal{F}_g^{(0), \rm NLP} &= -f_V^\perp m_V^2 m_Q \,  e^{-\frac{m_Q^2}{M^2}}\, \psi_3^\parallel(1/2)\,,\nn
	\mathcal{F}_g^{(0), \rm NNLP} &= \frac{1}{4}f_V^\parallel m_V^3 \,  e^{-\frac{m_Q^2}{M^2}}\, \Big[\phi_2^\parallel(1/2) +\Big(1+\frac{m_Q^2}{M^2} \Big) \big( \phi_4^\parallel(1/2) +8\,\psi_4^{\parallel,(2)}(1/2)\big) \Big]\,, \nn
	\mathcal{F}_g^{(1), \rm LP} &= f_V^\parallel m_V \int_{2m_Q^2}^{2s_0} ds\, e^{-\frac{s}{2M^2}}\, f_g\Big(\frac{s}{m_Q^2}-2\Big)\,.
	\label{result-Fg}
\end{align}
Similarly, the corresponding invariant functions $\mathcal{F}_f$ for the magnetic channel are given by
\begin{align}
	\mathcal{F}_f^{(0), \rm LP} &= f_V^\perp m_Q M^2\big(e^{-\frac{m_Q^2}{M^2}} -e^{-\frac{s_0}{M^2}} \big) \phi_2^\perp(1/2) \,, \nn
	\mathcal{F}_f^{(0), \rm NLP} &= \frac{1}{4}f_V^\parallel m_V \, \Big[2M^2 \big(e^{-\frac{m_Q^2}{M^2}} -e^{-\frac{s_0}{M^2}} \big) \big(\phi_3^\perp(1/2)+2\bar{\Phi}_3'(1/2)\big)+m_Q^2  e^{-\frac{m_Q^2}{M^2}}\, \psi_3^\perp(1/2)  \Big]\,,  \nn
	\mathcal{F}_f^{(0), \rm NNLP} &= -\frac{1}{4}f_V^\perp m_Q m_V^2 \,  e^{-\frac{m_Q^2}{M^2}}\, \Big[\phi_2^\perp(1/2) +2 \psi_4^{\perp,(1)}(1/2) +\frac{m_Q^2}{M^2} \phi_4^\perp(1/2)  \Big]\,, \nn 
	\mathcal{F}_f^{(1), \rm LP} &= -f_V^\perp m_Q \int_{2m_Q^2}^{2s_0} ds\, e^{-\frac{s}{2M^2}}\, f_f\Big(\frac{s}{m_Q^2}-2\Big)\,,
	\label{result-Ff}
\end{align}
where the integrated NLO spectral functions $f_{g(f)}(\sigma)$ read
\begin{align}
	f_g(\sigma) &=3 (3L +1) \delta(\sigma-0^+)  +3\theta(\sigma) \Big[-2{\rm Li}_2\Big(-\frac{\sigma}{2} \Big)+{\rm Li}_2\Big(-\frac{\sigma}{\sigma+2}\Big)\nn
	&\quad -{\rm Li}_2\Big(\frac{\sigma}{\sigma+2}\Big)
	- \ln\frac{\sigma+2}{2} \ln\frac{\sigma}{2} +\frac{\sigma^3+10\sigma^2+32\sigma+24}{2(\sigma+2)^3} \ln\frac{\sigma}{2} -\frac{1}{2}\ln\frac{\sigma+2}{2}\nn
	&\quad -\frac{2(\sigma+1)(\sigma+4)}{(\sigma+2)^3}\ln(\sigma+1) + \frac{\sigma^2+4\sigma+5}{(\sigma+2)^2} -\frac{\pi^2}{6}\Big] \,, \nn
	f_f(\sigma) &=3 (3L +4) \delta(\sigma-0^+) +3\theta(\sigma) \Big[-2{\rm Li}_2\Big(-\frac{\sigma}{2} \Big)+{\rm Li}_2\Big(-\frac{\sigma}{\sigma+2}\Big)\nn
	&\quad -{\rm Li}_2\Big(\frac{\sigma}{\sigma+2}\Big)
	- \ln\frac{\sigma+2}{2} \ln\frac{\sigma}{2} +\frac{(\sigma+4)(3\sigma+4)}{2(\sigma+2)^2} \ln\frac{\sigma}{2} -\frac{1}{2}\ln\frac{\sigma+2}{2}\nn
	&\quad -\frac{2(\sigma+1)}{(\sigma+2)^2}\ln(\sigma+1) + \frac{1}{\sigma+2}-\frac{\pi^2}{6}+\frac{1}{4}-L\Big] \,.
\end{align}
It follows that the parameter $f_{H^*H^*V}$ is power-enhanced relative to $g_{H^*H^*V}$, dynamically compensating for the kinematic suppression present in Eq.~(\ref{kinematic}). This scaling compensation, anticipated in Eq.~(\ref{eq:HQSS_breaking}), will be analytically verified in the heavy quark limit in Section~\ref{sec-4}.

\section{Numerical analysis}
\label{sec-4}

%\subsection{Input parameters}

To perform the numerical study of the established LCSR in Eq.~(\ref{result}) for both the charge and magnetic $H^*H^*V$ strong couplings, we first specify the numerical input values for the relevant QCD and hadronic parameters, which are collected in Table~\ref{tab_1}. Consistent with our theoretical setup, we utilize the running heavy quark masses evaluated in the $\overline{\text{MS}}$ scheme. To achieve high numerical precision and minimize lattice systematic errors, we determine the heavy vector meson decay constants by combining the  LQCD results for the pseudoscalar decay constants~\cite{FLAG:2024oxs} with the vector-to-pseudoscalar mass and decay constant ratios~\cite{Lubicz:2017asp}. To reflect the typical virtuality of the propagating heavy quark in the correlation function, we vary the factorization scale $\mu$ within the range of $[1.0,\,3.0]$~GeV around a central value of $1.5$~GeV for the charm sector, and within $[m_b/2,\,2m_b]$ around a central value of $\overline{m}_b(\overline{m}_b)$ for the bottom sector. The optimal working windows for the Borel parameters $M^2$ and the effective continuum thresholds $s_0$ are determined by demanding both ground-state dominance and  OPE convergence, which are consistent with previous LCSR analyses~\cite{Li:2020rcg}.
\begin{table}[h]
	\centering 
	\renewcommand{\arraystretch}{1.0} 
	\setlength{\tabcolsep}{4pt} 
	\resizebox{\textwidth}{!}{ 
		\begin{tabular}{|ll|ll|} 
			\hline
			Parameter & Value & Parameter & Value \\
			\hline
			\multicolumn{2}{|l|}{\textbf{Quark Masses} \cite{ParticleDataGroup:2026aaa}} & \multicolumn{2}{l|}{\textbf{Decay Constants} \cite{FLAG:2024oxs,Lubicz:2017asp}} \\
			$\overline{m}_u\, (2 {\rm GeV})$ & $2.16\pm 0.07$ MeV & $f_{D^*}$ & $228.5\pm 7.7$ MeV \\
			$\overline{m}_d\, (2 {\rm GeV})$ & $4.70\pm 0.07$ MeV & $f_{D_s^*}$ & $271.6\pm 5.0$ MeV \\
			$\overline{m}_s\, (2 {\rm GeV})$ & $92.9\pm 0.7$ MeV & $f_{B^*}$ & $182.0\pm 4.4$ MeV \\
			$\overline{m}_c(\overline{m}_c)$ & $1.273\pm0.005\, {\rm GeV}$ & $f_{B_s^*}$ & $224.3\pm 2.6$ MeV \\
			$\overline{m}_b(\overline{m}_b)$ & $4.186\pm0.006 $ GeV & & \\
			\multicolumn{2}{|l|}{\textbf{Charmed Meson Sum Rules} \cite{Li:2020rcg}} & \multicolumn{2}{l|}{\textbf{Bottom Meson Sum Rules} \cite{Li:2020rcg}} \\
			$\mu$ & $1.5_{-0.5}^{+1.5} $ GeV & $\mu$ & ${m_b}_{-m_b/2}^{+m_b} $ \\
			$\{s_0, M^2\}$ ($D^*D^*V$) & $\{6.0\pm0.5, 4.5\pm 1.0\} \, {\rm GeV}^2$ & $\{s_0, M^2\}$ ($B^*B^*V$) & $\{34.0\pm1.0, 18.0\pm 3.0\}\, {\rm GeV}^2 $ \\
			$\{s_0, M^2\}$ ($D_s^*D_{(s)}^*V$) & $\{7.0\pm0.5, 4.5\pm 1.0\} \, {\rm GeV}^2$ & $\{s_0, M^2\}$ ($B_s^*B_{(s)}^*V$) & $\{36.0\pm1.0, 18.0\pm 3.0\}\, {\rm GeV}^2 $ \\
			\hline 
		\end{tabular} 
	}
	\caption{Summary of numerical input parameters, including quark masses, $H^*$ meson decay constants, factorization scales $\mu$, Borel masses $M^2$, and duality thresholds $s_0$ employed in the LCSR.}
	\label{tab_1}
\end{table}
The fundamental hadronic parameters governing the light vector meson DAs, along with their renormalization group evolution equations, are detailed in Appendix~\ref{appendix-A}. %Unlike the exactly scale-independent longitudinal decay constant $f_V^\parallel$, the remaining hadronic parameters (including the transverse decay constants, twist-two Gegenbauer moments, and higher-twist parameters) are evolved to match the factorization scale $\mu$ of the hard scattering kernel.
%\subsection{OPE convergence and sum rule stability}

To demonstrate the theoretical reliability of the established LCSR, we first examine the convergence of the OPE. In Table~\ref{tab_OPE}, we present the numerical breakdown of the products $g_{H^*H^*V} f_{H_1^*} f_{H_2^*}$ and $f_{H^*H^*V} f_{H_1^*} f_{H_2^*}$ into their individual components from different twists and perturbative orders at the central input values. As anticipated, the LP contribution at LO dominates the sum rules. The power-suppressed NLP corrections offer a non-negligible positive enhancement, which is slightly more pronounced than the NLO perturbative adjustments of LP. Crucially, the NNLP terms yield an insignificantly small contribution for both channels, thereby validating the truncation of the higher-twist series and ensuring the convergence of the OPE within our theoretical framework.
\begin{table}[h]
	\setlength{\tabcolsep}{8pt}
	\centering 
	\begin{tabular}{|l|ccccc|} 
		\hline 
		Coupling Channel & LP LO & LP NLO & NLP & NNLP & Total \\     
		\hline
		$g_{D^*D^*\rho}\, f_{D^*}^2$ &$0.125$ & $0.016$ & $0.030$ &$-0.012$ &$0.159$  \\
		$g_{D^*D^*\omega}\, f_{D^*}^2$  &$0.085$ & $0.011$ & $0.020$ &$-0.008$ &$0.107$  \\
		$g_{D^*_sD^*K^*}\, f_{D_s^*}\, f_{D^*}$  &$0.156$ & $0.018$ & $0.038$ &$-0.018$ &$0.194$  \\
		$g_{D_s^*D_s^*\phi}\, f_{D_s^*}^2$  &$0.180$ & $0.024$ & $0.054$ &$-0.028$ &$0.231$  \\
		\hline
		$g_{B^*B^*\rho}\, f_{B^*}^2$ &$0.073$ & $0.015$ & $0.016$ &$-0.002$ &$0.101$  \\
		$g_{B^*B^*\omega}\, f_{B^*}^2$ &$0.049$ & $0.010$ & $0.011$ &$-0.001$ &$0.068$  \\
		$g_{B^*_sB^*K^*}\, f_{B_s^*}\, f_{B^*}$  &$0.088$ & $0.017$ & $0.021$ &$-0.003$ &$0.122$  \\
		$g_{B_s^*B_s^*\phi} \, f_{B_s^*}^2$  &$0.107$ & $0.023$ & $0.031$ &$-0.005$ &$0.154 $  \\
		\hline
		\hline
		$f_{D^*D^*\rho}\, f_{D^*}^2$ &$0.075$ & $0.013$ & $0.023$ &$-0.005$ &$0.106$  \\
		$f_{D^*D^*\omega}\, f_{D^*}^2$  &$0.049$ & $0.009$ & $0.015$ &$-0.004$ &$0.069$  \\
		$f_{D^*_sD^*K^*}\, f_{D_s^*}\, f_{D^*}$  &$0.088$ & $0.013$ & $0.030$ &$-0.007$ &$0.123$  \\
		$f_{D_s^*D_s^*\phi}\, f_{D_s^*}^2$  &$0.099$ & $0.016$ & $0.039$ &$-0.011$ &$0.143$  \\
		\hline
		$f_{B^*B^*\rho}\, f_{B^*}^2$ &$0.150$ & $0.014$ & $0.025$ &$-0.004$ &$0.184$  \\
		$f_{B^*B^*\omega}\, f_{B^*}^2$ &$0.098$ & $0.009$ & $0.016$ &$-0.003$ &$0.121$  \\
		$f_{B^*_sB^*K^*}\, f_{B_s^*}\, f_{B^*}$  &$0.169$ & $0.015$ & $0.030$ &$-0.006$ &$0.207$  \\
		$f_{B_s^*B_s^*\phi} \, f_{B_s^*}^2$  &$0.195$ & $0.019$ & $0.038$ &$-0.009$ &$0.243 $  \\
		\hline
		\hline
	\end{tabular} 
	\caption{Numerical breakdown of the products $g_{H^*H^*V} f_{H_1^*} f_{H_2^*}$ (in units of ${\rm GeV}^2$) at their central input values, detailing contributions from individual OPE components.}
	\label{tab_OPE}
\end{table}

\begin{figure}[htpb]
	\begin{center}
		\includegraphics[width=1.0 \columnwidth]{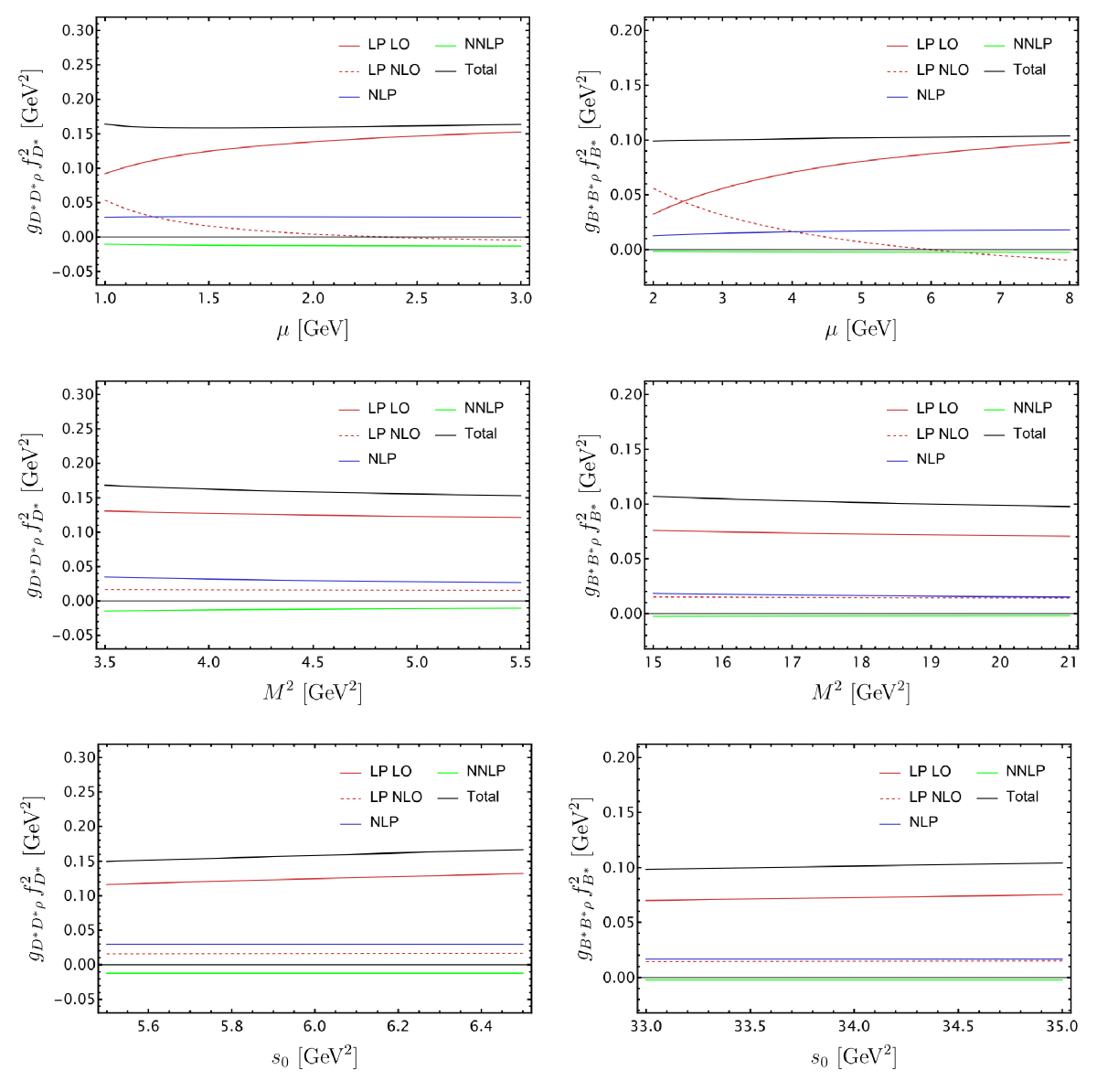} \\
		\caption{Dependence of the LCSR predictions for $g_{D^{*}D^{*}\rho}\, f_{D^{*}}^2$ (left column) and $g_{B^{*}B^{*}\rho}\,f_{B^{*}}^2$ (right column) on the factorization scale $\mu$, Borel mass $M^2$, and effective continuum threshold $s_0$, with all other input parameters fixed at their central values.}
		\label{depedence-g}
	\end{center}
\end{figure}

\begin{figure}[htpb]
	\begin{center}
		\includegraphics[width=1.0 \columnwidth]{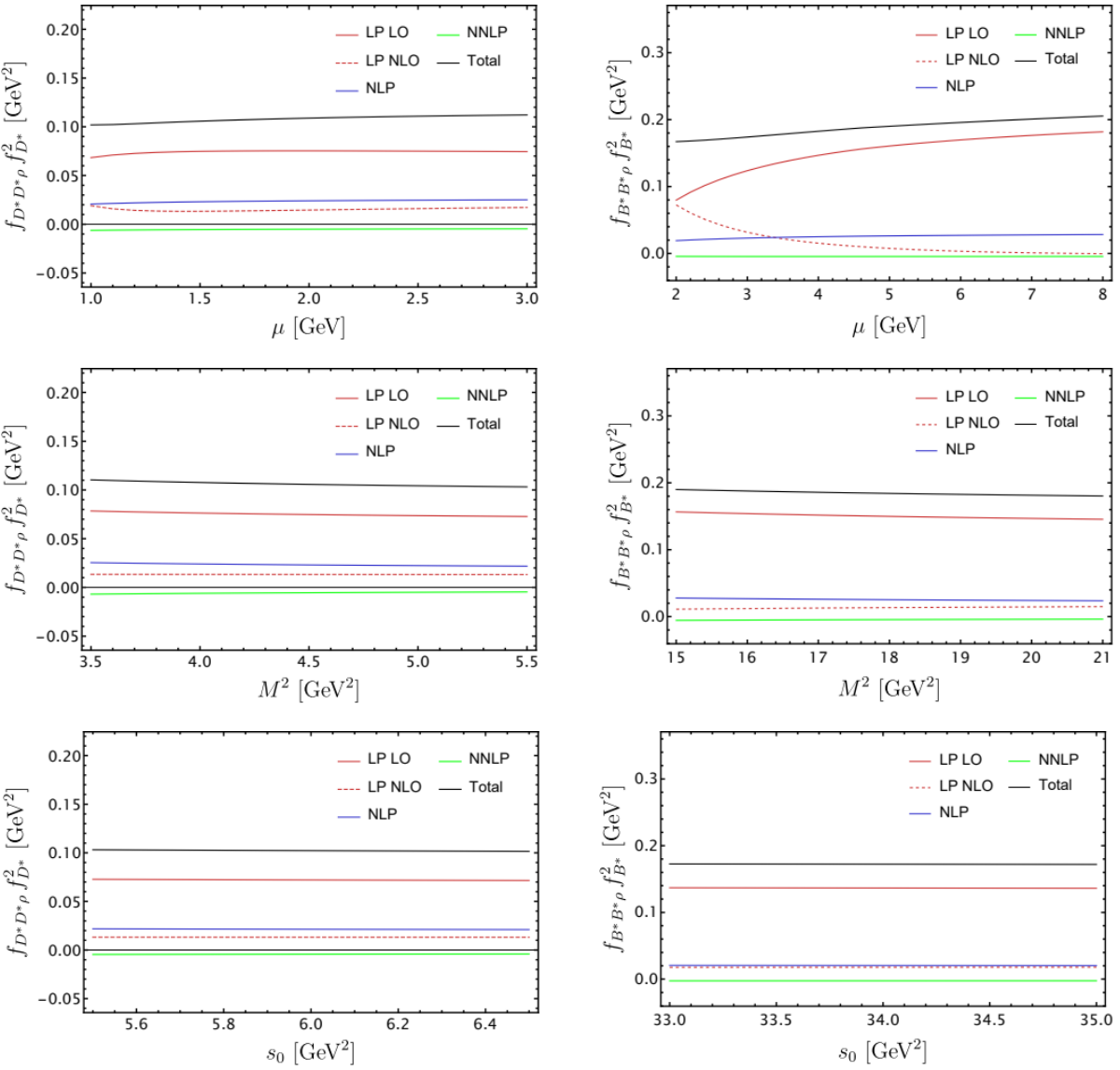} \\
		\caption{Same as Fig.~\ref{depedence-g}, but for the magnetic couplings $f_{D^{*}D^{*}\rho}\, f_{D^{*}}^2$ (left column) and $f_{B^{*}B^{*}\rho}\,f_{B^{*}}^2$ (right column).}
		\label{depedence-f}
	\end{center}
\end{figure}

Furthermore, we investigate the stability of the extracted couplings with respect to variations in the factorization scale $\mu$, the Borel parameter $M^2$, and the effective continuum threshold $s_0$. In Figs.~\ref{depedence-g} and \ref{depedence-f}, we illustrate this parameter dependence for the charge and magnetic couplings, respectively, taking the $D^*D^*\rho$ (left column) and $B^*B^*\rho$ (right column) channels as representative examples. In the top panels, the inclusion of the LP NLO correction effectively cancels the scale dependence of the LO contribution, yielding stable total predictions across a wide range of $\mu$. As shown in the middle and bottom panels, the stable Borel plateaus justify the quark-hadron duality ansatz, while the mild sensitivity to the continuum threshold $s_0$ is incorporated into the theoretical error budget.

%\subsection{Final results and physical implications}

After factoring out the heavy vector meson decay constants, we obtain our final predictions for the strong charge and magnetic couplings, as summarized in Tables~\ref{final_result_g} and \ref{final_result_f}. The total uncertainties are evaluated by adding the individual parameter variations in quadrature. Here, $\Delta\alpha$ denotes the uncertainty induced by the geometric shape of the two-dimensional duality region. Varying the boundary parameterization ($\alpha=1/2, 1, 2$) yields a minor deviation of approximately $5\%$ on the final extractions, firmly validating the robustness of our duality ansatz. A detailed breakdown of the error budget reveals that the dominant sources of theoretical uncertainty originate from the input variances of the heavy meson decay constants $f_{H^*}$ and the twist-two Gegenbauer moments of the light vector mesons.
\begin{table}[htpb]
	\setlength{\tabcolsep}{8pt}
	\centering 
	\begin{tabular}{|c|ccccccccc|} 
		\hline 
		Coupling &Central& $\Delta f_{H^*}$ & $\Delta \mu$ &  $\Delta M^2$&  $\Delta s_0$ & $\Delta f_V^\parallel$ & $\Delta a_2^\parallel$ & $\Delta \alpha$& Total \\
		\hline
		$g_{D^*D^*\rho}$ & $3.04$ & $_{-0.19}^{+0.22}$ & $_{-0.00}^{+0.10}$& $_{-0.10}^{+0.19}$ & $_{-0.17}^{+0.15}$  & $\pm 0.06$ & $\pm 0.24$ & $_{-0.16}^{+0.09}$ & $_{-0.41}^{+0.44}$ \\
		$g_{D^*D^*\omega}$ & $2.05$ & $_{-0.13}^{+0.15}$ & $_{-0.00}^{+0.07}$& $_{-0.07}^{+0.13}$ & $_{-0.11}^{+0.10}$  & $\pm 0.10$ & $\pm 0.27$ & $_{-0.10}^{+0.06}$ & $_{-0.36}^{+0.38}$ \\
		$g_{D^*_sD^*K^*}$ & $3.13$ & $_{-0.11}^{+0.12}$ & $_{-0.00}^{+0.11}$& $_{-0.09}^{+0.18}$ & $_{-0.13}^{+0.11}$  & $\pm 0.08$ & $\pm 0.31$ & $_{-0.15}^{+0.08}$ & $_{-0.41}^{+0.44}$ \\
		$g_{D^*_sD_s^*\phi}$ & $3.13$ & $_{-0.11}^{+0.12}$ & $_{-0.00}^{+0.12}$& $_{-0.12}^{+0.23}$ & $_{-0.12}^{+0.11}$  & $\pm 0.04$ & $\pm 0.29$ & $_{-0.16}^{+0.09}$ & $_{-0.40}^{+0.45}$ \\
		\hline
		$g_{B^*B^*\rho}$ & $3.06$ & $_{-0.14}^{+0.15}$ & $_{-0.06}^{+0.08}$& $_{-0.11}^{+0.17}$ & $\pm0.09$  & $\pm 0.06$ & $\pm 0.16$ & $_{-0.07}^{+0.04}$ & $_{-0.28}^{+0.31}$ \\
		$g_{B^*B^*\omega}$ & $2.05$ & $\pm0.10$ & $_{-0.05}^{+0.06}$& $_{-0.07}^{+0.11}$ & $\pm0.06$  & $\pm 0.10$ & $\pm 0.18$ & $_{-0.05}^{+0.03}$ & $_{-0.26}^{+0.27}$ \\
		$g_{B^*_sB^*K^*}$ & $3.00$ & $\pm0.08$ & $_{-0.06}^{+0.08}$& $_{-0.11}^{+0.16}$ & $_{-0.08}^{+0.07}$  & $\pm 0.08$ & $\pm 0.19$ & $_{-0.07}^{+0.04}$ & $_{-0.28}^{+0.30}$ \\
		$g_{B^*_sB_s^*\phi}$ & $3.08$ & $\pm0.07$ & $_{-0.02}^{+0.07}$& $_{-0.12}^{+0.19}$ & $_{-0.08}^{+0.07}$  & $\pm 0.04$ & $\pm 0.18$ & $_{-0.07}^{+0.04}$ & $_{-0.26}^{+0.29}$ \\
		\hline
	\end{tabular} 
	\caption{Detailed breakdown of the theoretical uncertainties for the extracted strong couplings $g_{H^{*}H^{*}V}$. The total uncertainties are obtained by adding the individual variations in quadrature.}
	\label{final_result_g}
\end{table}

\begin{table}[h]
	\setlength{\tabcolsep}{8pt}
	\centering 
	\begin{tabular}{|c|ccccccccc|} 
		\hline 
		Coupling &Central& $\Delta f_{H^*}$ & $\Delta \mu$ &  $\Delta M^2$&  $\Delta s_0$ & $\Delta f_V^\perp$ & $\Delta a_2^\perp$ & $\Delta \alpha$& Total \\
		\hline
		$f_{D^*D^*\rho}$ & $2.03$ & $_{-0.13}^{+0.14}$ & $_{-0.07}^{+0.12}$& $_{-0.05}^{+0.09}$ & $_{-0.12}^{+0.11}$  & $\pm 0.07$ & $\pm 0.12$ & $_{-0.10}^{+0.06}$ & $_{-0.26}^{+0.28}$ \\
		$f_{D^*D^*\omega}$ & $1.33$ & $\pm0.09$ & $_{-0.05}^{+0.08}$& $_{-0.03}^{+0.06}$ & $_{-0.08}^{+0.07}$  & $\pm 0.13$ & $\pm 0.15$ & $_{-0.06}^{+0.04}$ & $\pm0.25$ \\
		$f_{D^*_sD^*K^*}$ & $1.99$ & $_{-0.07}^{+0.08}$ & $_{-0.06}^{+0.09}$& $_{-0.04}^{+0.07}$ & $_{-0.09}^{+0.07}$  & $\pm 0.06$ & $\pm 0.15$ & $_{-0.08}^{+0.05}$ & $_{-0.22}^{+0.24}$ \\
		$f_{D^*_sD_s^*\phi}$ & $1.94$ & $\pm0.07$ & $_{-0.09}^{+0.12}$& $_{-0.05}^{+0.09}$ & $_{-0.08}^{+0.07}$  & $\pm 0.03$ & $\pm 0.13$ & $_{-0.08}^{+0.05}$ & $_{-0.22}^{+0.24}$ \\
		\hline
		$f_{B^*B^*\rho}$ & $5.56$ & $_{-0.26}^{+0.28}$ & $_{-0.50}^{+0.69}$& $_{-0.12}^{+0.17}$ & $_{-0.22}^{+0.21}$  & $\pm 0.21$ & $\pm 0.27$ & $_{-0.16}^{+0.09}$ & $_{-0.72}^{+0.86}$ \\
		$f_{B^*B^*\omega}$ & $3.64$ & $_{-0.17}^{+0.18}$ & $_{-0.33}^{+0.45}$& $_{-0.08}^{+0.11}$ & $_{-0.14}^{+0.13}$  & $\pm 0.39$ & $\pm 0.35$ & $_{-0.10}^{+0.06}$ & $_{-0.67}^{+0.74}$ \\
		$f_{B^*_sB^*K^*}$ & $5.06$ & $_{-0.13}^{+0.14}$ & $_{-0.44}^{+0.57}$& $_{-0.10}^{+0.15}$ & $_{-0.16}^{+0.15}$  & $\pm 0.16$ & $\pm 0.31$ & $_{-0.14}^{+0.08}$ & $_{-0.63}^{+0.72}$ \\
		$f_{B^*_sB_s^*\phi}$ & $4.83$ & $\pm0.11$ & $_{-0.41}^{+0.57}$& $_{-0.11}^{+0.16}$ & $_{-0.16}^{+0.15}$  & $\pm 0.09$ & $\pm 0.27$ & $_{-0.13}^{+0.07}$ & $_{-0.57}^{+0.68}$ \\
		\hline
	\end{tabular} 
	\caption{Same as Table~\ref{final_result_g}, but for the magnetic couplings $f_{H^{*}H^{*}V}$.}
	\label{final_result_f}
\end{table}

A comprehensive comparison of our predictions for $g_{H^*H^*V}$ with existing literature is summarized in Table~\ref{compare}. Our LCSR results are  distinct from traditional QCDSR extractions as we directly evaluate the on-shell parameters without introducing additional shape-dependent extrapolations. When comparing our absolute value for $g_{D^*D^*\rho}$ with the LQCD estimate \cite{Can:2012tx}, a noticeable tension is observed. The lattice result is obtained by extrapolating the $D^*$ electromagnetic form factors to the meson pole, which may introduce potential systematic uncertainties. Translating their specific value of $g_{D^*D^*\rho}$ to the heavy quark limit would yield an anomalously large universal coupling $\beta \sim 1.4$.
\begin{table}[htpb]
	\setlength{\tabcolsep}{8pt}
	\centering 
	\begin{tabular}{|c|cccc|} 
		\hline 
		Method  & $g_{D^*D^*\rho}$ & $g_{D^*D^*\omega}$ & $g_{D_s^*D^*K^*}$ &$g_{D_s^*D_s^*\phi}$  \\
		\hline
		this work & $3.04_{-0.41}^{+0.44}$ & $2.05_{-0.36}^{+0.38}$ & $3.13_{-0.41}^{+0.44}$ & $3.13_{-0.40}^{+0.45}$\\
		%\hline
		LCSR~\cite{Aliev:2021cjt} &$1.8\pm 0.4$ & $1.5\pm 0.3$ & $1.7\pm 0.4$ & $1.5\pm 0.4$   \\
		LCSR~\cite{Wang:2007ci}  & $2.6\pm0.7$ & $-$ &$-$ &$-$\\
		QCDSR &$6.6\pm 0.3$~\cite{Bracco:2011pg}  & $-$ & $4.95\pm0.64$~\cite{Janbazi:2018bgv} & $7.52\pm 1.70$~\cite{Khosravi:2013ad}  \\
		LQCD~\cite{Can:2012tx} & $5.94\pm0.56$ &$-$ &$-$ &$-$\\ 
		\hline
		\hline
		Method  & $g_{B^*B^*\rho}$ & $g_{B^*B^*\omega}$ & $g_{B_s^*B^*K^*}$ &$g_{B_s^*B_s^*\phi}$ \\
		\hline
		this work  & $3.06_{-0.28}^{+0.31}$ & $2.05_{-0.26}^{+0.27}$ & $3.00_{-0.28}^{+0.30}$ & $3.08_{-0.26}^{+0.29}$ \\
		%\hline
		LCSR~\cite{Aliev:2021cjt} &$1.1\pm 0.2$ & $0.9\pm 0.2$ & $1.1\pm 0.2$ & $1.0\pm 0.2$ \\
		LCSR~\cite{Li:2007dv}  & $2.66$ & $-$ &$-$ &$-$\\
		QCDSR~\cite{Cui:2012wk}  & $2.53\pm0.83$ & $-$ &$-$ &$-$\\
		\hline
	\end{tabular} 
	\caption{Comparison of the extracted strong couplings $g_{D^*D^*V}$ and $g_{B^*B^*V}$ with previous theoretical determinations.}
	\label{compare}
\end{table}
Furthermore, for the magnetic couplings, previous leading-order LCSR~\cite{Li:2007dv} and three-point QCDSR~\cite{Cui:2012wk} calculations correspond to $f_{B^*B^*\rho} \approx 6.17$ and $7.08 \pm 1.81$, respectively, after accounting for an isospin and mass convention factor of $\sqrt{2}m_{B^*}$ to match our dimensionless definition. Our updated prediction ($f_{B^*B^*\rho} = 5.56_{-0.72}^{+0.86}$) is noticeably smaller. Although the NLO and higher-twist corrections provide a moderate positive enhancement, this overall reduction is driven by recent updates to non-perturbative inputs, particularly the heavy meson decay constants and transverse DAs.

With these updated results, we proceed to investigate the HQSS breaking effects. To demonstrate the theoretical consistency of our framework and elucidate the origin of this symmetry, it is highly instructive to examine the asymptotic behavior of the derived LCSRs in the heavy quark limit. By applying the standard heavy-quark scaling relations to the leading-order amplitudes~\cite{Khodjamirian:2020mlb}:
\begin{align}
	m_{H^*} = m_Q + \bar{\Lambda}\,, \quad M^2 = 2m_Q \tau\,, \quad s_0 = m_Q^2 + 2m_Q \omega_0\,, \quad f_{H^*} = \frac{\hat{f}}{\sqrt{m_Q}}\,,
\end{align}
where $\bar{\Lambda}$ and $\hat{f}$ are the binding energy and the static decay constant of the heavy meson in Heavy Quark Effective Theory (HQET), respectively. Retaining the leading power in $1/m_Q$, the LO sum rules analytically reduce to
\begin{align}
	g_{H^*H^*V} &\simeq \frac{m_V}{\hat{f}^2} e^{\bar{\Lambda}/\tau} \Big[ \tau f_V^\parallel \left(1 - e^{-\frac{\omega_0}{\tau}}\right) \phi_2^\parallel (1/2) + \frac{f_V^\perp m_V}{2} \psi_3^\parallel(1/2)  \nn
	&\quad  - \frac{f_V^\parallel m_V^2}{16\tau} \Big( \phi_4^\parallel(1/2) + 8\psi_4^{\parallel,(2)}(1/2) \Big) \Big] \,, \nn
	f_{H^*H^*V} &\simeq \frac{m_Q}{2\hat{f}^2} e^{\bar{\Lambda}/\tau} \Big[ \tau f_V^\perp \left(1 - e^{-\frac{\omega_0}{\tau}}\right) \phi_2^\perp (1/2) + \frac{f_V^\parallel m_V}{8} \psi_3^\perp(1/2) 
	 - \frac{f_V^\perp m_V^2}{16\tau} \phi_4^\perp(1/2) \Big]\,. \label{eq:HQET_limit_f}
\end{align}
This confirms that our LCSR framework dynamically preserves the HQSS expectations outlined in Eq.~(\ref{eq:HQSS_breaking}). To properly accommodate the NLO corrections, we extract the universal static coupling directly via a numerical fit of the full finite-mass sum rules. Following the parameterization in Eq.~(\ref{eq:HQSS_breaking}), we isolate the $\mathcal{O}(1/m_{H^*})$ power corrections from the static limit. Utilizing the $H^*H^*\rho$ channels  and adopting the value $f_\pi = 130.2$ MeV \cite{FLAG:2024oxs}, we extract the universal static coupling and the mass-correction parameter
\begin{align}
	\beta &= 0.73 \pm 0.13 \,, \hspace{1.7cm} \delta_a = -0.02 \pm 0.54 \text{ GeV}\,, \nn
	\hat{\lambda} &= 0.23 \pm 0.06 \text{ GeV}^{-1}\,, \quad \delta_b = -0.13 \pm 0.58 \text{ GeV}\,.
	\label{static}
\end{align}
The extracted value of $\beta$ is  consistent with the consensus range of $0.7 \sim 0.9$ favored by various non-perturbative approaches, such as QCDSR and constituent quark models. Meanwhile, for the magnetic coupling parameter, we obtain $\hat{\lambda} = 0.23 \pm 0.06 \text{ GeV}^{-1}$, which is moderately smaller than the typical range of $0.3 \sim 0.6 \text{ GeV}^{-1}$ estimated from early leading-order LCSR and vector-meson dominance models~\cite{Casalbuoni:1996pg,Li:2007dv,Isola:2003fh}.  Unlike previous LO approximations, our framework  incorporates the NLO radiative corrections and power suppressed higher-twist contributions. In particular, the exact inclusion of the transverse DAs (such as $\phi_2^\perp$ and $\psi_3^\perp$) significantly rectifies the extraction of the spin-flip parameter $\hat{\lambda}$.

Finally, we address the $SU(3)$ flavor symmetry breaking effects. To quantify this breaking, we define the relative deviation parameter, taking the charge couplings in the charm sector as an example,
\begin{align}
	\delta_K=\frac{g_{D_s^*D^*K^*} }{g_{D^*D^*\rho}}-1 \,.
\end{align}
Evaluated at the central input values, this $SU(3)$ symmetry departure is small, yielding $\delta_K \approx 3\%$. However, due to a dynamical enhancement in the LO coefficient of the second Gegenbauer moment, the input uncertainties from $a_2^{\parallel, \rho}$ and $a_2^{\parallel, K^*}$ are severely magnified, translating into relative errors of approximately $8\%$ and $10\%$ for the individual couplings in the charm sector which yields a total relative uncertainty of about $13\%$ for LP contribution. 

To preserve physical consistency, we vary the factorization scale $\mu$, Borel mass $M^2$, and threshold $s_0$ concertedly for both channels, while treating other inputs as completely uncorrelated. This evaluation yields $\delta_K = (3_{-16}^{+18})\%$. This wide error band  demonstrates that the subtle $SU(3)$ breaking deviations are largely submerged by current theoretical uncertainties, which are overwhelmingly dominated by the precision limits of the vector meson DAs. For the magnetic couplings $f_{H^*H^*V}$, the inherently small symmetry departure (e.g., $f_{D_s^*D^*K^*}/f_{D^*D^*\rho}-1 \approx -2\%$) is similarly washed out by the uncertainties of the transverse DA parameters.
Therefore, current uncertainties in the vector meson DAs preclude a precise extraction of the $SU(3)$ flavor symmetry breaking.

\section{Conclusion and outlook}
\label{sec-5}

In this work, we have presented a  derivation of both the strong charge couplings $g_{H^*H^*V}$ and magnetic couplings $f_{H^*H^*V}$ ($H \in \{D, B\}$ and $V \in \{\rho, \omega, K^*, \phi\}$) within the framework of LCSR. To achieve high theoretical accuracy, our analysis goes beyond the LO approximation by incorporating NLO perturbative QCD corrections to the twist-two DAs. We further include power suppressed contributions up to NNLP accuracy, adopting asymptotic forms of DAs for the singular terms to ensure the cancellation of anomalous boundary contributions.
From the analytical LCSR expressions, we demonstrated that the magnetic couplings are power-enhanced relative to the charge couplings, which counterbalances the kinematic momentum suppression. Taking the heavy quark limit further confirms that this dynamic compensation strictly preserves the underlying HQSS. 

Our numerical analysis reveals an approximate degeneracy for the charge couplings between the charm and bottom sectors, yielding $g_{D^*D^*\rho} = 3.04_{-0.41}^{+0.44}$ and $g_{B^*B^*\rho} = 3.06_{-0.28}^{+0.31}$. Conversely, the magnetic couplings exhibit a clear heavy-quark mass hierarchy, with $f_{D^*D^*\rho} = 2.03_{-0.26}^{+0.28}$ and $f_{B^*B^*\rho} = 5.56_{-0.72}^{+0.86}$. We note that although the magnetic couplings are parametrically power-enhanced relative to the charge couplings, the numerical value of $f_{D^*D^*\rho}$ is smaller than $g_{D^*D^*\rho}$ in the charm sector. This is a direct consequence of the specific numerical coefficients in the sum rules. The transverse decay constant is naturally smaller ($f_V^\perp < f_V^\parallel$), and the magnetic channel carries a smaller overall prefactor ($1/4$ versus $1/2$). In the charm sector, the moderate heavy quark mass ($m_c \approx 1.3 \text{ GeV}$) is insufficient to overcome these numerical suppressions. Conversely, in the bottom sector, the large heavy quark mass ($m_b \approx 4.2 \text{ GeV}$) dominates the scaling behavior, correctly yielding the numerically enhanced hierarchy $f_{B^*B^*\rho} > g_{B^*B^*\rho}$.

By matching our sum rules with HM$\chi$PT, we extracted the universal static couplings $\beta = 0.73 \pm 0.13$ and $\hat{\lambda} = 0.23 \pm 0.06 \text{ GeV}^{-1}$, which are structurally consistent with standard phenomenological expectations. Furthermore, we investigated the $SU(3)$ flavor symmetry breaking effects across the different channels. Although the central values of the relative $SU(3)$ departures are small ($\sim 3\%$ for the charge couplings and $\sim -2\%$ for the magnetic ones), our comprehensive error analysis shows that these tiny flavor-symmetry departures are currently overwhelmed by theoretical uncertainties originating from the light vector meson DAs.

From a phenomenological perspective, extracting these strong couplings with high precision is crucial for modeling meson-exchange potentials and facilitating the search for exotic hadronic molecules, such as the $T_{cc}^*$ state, near heavy-meson thresholds. By evaluating both channels at the NLO+NNLP accuracy, our unified framework provides the complete set of non-perturbative inputs required for coupled-channel dynamics. The precise determination of both the dominant spin-conserving charge couplings and the complementary spin-flip magnetic couplings establishes a robust theoretical foundation for quantitatively understanding the intricate binding mechanisms in exotic heavy tetraquarks. Enhancing the predictive accuracy of these sum rules will  rely on improved determinations of the vector meson DAs from upcoming theoretical and experimental advances

\section*{Acknowledgments}
This work was supported in part by the National Natural Science Foundation of China under Grant No. 12105112.

\appendix

\section{The vector meson distribution amplitudes}
\label{appendix-A}

The definitions of the two-particle DAs up to twist-5 accuracy  are given as follows~\cite{Bharucha:2015bzk}:
\begin{align}
	&\langle V(p,\epsilon) | \bar{q}_1(x) \gamma_\mu q_2(0) |0\rangle \nn
	&= f_V^\parallel m_V \int_0^1 du\, e^{iup\cdot x} \Biggl\{ p_\mu \frac{\epsilon^*\cdot x }{p \cdot x} \Big[ \phi_2^\parallel(u) -\phi_3^\perp(u) +\frac{m_V^2 x^2}{16}  (\phi_4^\parallel(u) -\phi_5^\perp(u)) \Big]   \nn
	&\quad +\epsilon_\mu^*\Big[ \phi_3^\perp(u) + \frac{m_V^2 x^2}{16} \phi_5^\perp(u) \Big]  - \frac{\epsilon^*\cdot x}{2(p\cdot x)^2} x_\mu m_V^2 \Big[ \psi_4^\parallel(u) -2\phi_3^\perp(u) +\phi_2^\parallel(u) \Big] \Biggr\} \nn
	&= f_V^\parallel m_V \int_0^1 du\, e^{iup\cdot x} \Biggl\{ (-i)p_\mu \,\epsilon^*\cdot x  \Big[ \hat{\phi}_2^\parallel(u) -\hat{\phi}_3^\perp(u) +\frac{m_V^2 x^2}{16}  (\hat{\phi}_4^\parallel(u) -\hat{\phi}_5^\perp(u)) \Big]   \nn
	&\quad +\epsilon_\mu^*\Big[ \phi_3^\perp(u) + \frac{m_V^2 x^2}{16} \phi_5^\perp(u) \Big]  + \frac{\epsilon^*\cdot x}{2} x_\mu m_V^2 \Big[ \hat{\hat{\psi}}_4^\parallel(u) -2\hat{\hat{\phi}}_3^\perp(u) +\hat{\hat{\phi}}_2^\parallel(u) \Big] \Biggr\} \nn
	&\equiv f_V^\parallel m_V \int_0^1 du\, e^{iup\cdot x} \Biggl\{ (-i)p_\mu \,\epsilon^*\cdot x  \Big[ \phi_2^{\parallel,(1)}(u) +\frac{m_V^2 x^2}{16}  \phi_4^{\parallel,(1)}(u)  \Big]   \nn
	&\quad +\epsilon_\mu^*\Big[ \phi_3^\perp(u) + \frac{m_V^2 x^2}{16} \phi_5^\perp(u) \Big]  + \frac{\epsilon^*\cdot x}{2} x_\mu m_V^2  \psi_4^{\parallel,(2)}(u)   \Biggr\}\,, \label{DAs1} \\
	&\langle V(p,\epsilon)|\bar{q}_1(x)\sigma_{\mu\nu}q_2(0)|0\rangle 
	=-i\,f_V^\perp  \int_0^1du\,e^{i\,u\,p\cdot x}\,\Bigg\{ \epsilon^*_{[\mu}\,p_{\nu]}  \Big[\phi_2^\perp(u)+\frac{m_V^2\,x^2}{16}\,\phi_4^\perp(u)\Big] \nonumber\\
	&\quad +p_{[\mu}\,x_{\nu]}\,\frac{\epsilon^*\cdot x}{(p\cdot x)^2}\,m_V^2 \Big[\phi_3^\parallel(u) -\frac{1}{2} \phi_2^\perp(u) -\frac{1}{2} \psi_4^\perp(u) \Big] 
	+\frac{\epsilon^*_{[\mu}\,x_{\nu]}}{2 p\cdot x}\, m_V^2 \big[\psi_4^\perp(u) -\phi_2^\perp(u)\big] \Bigg\}  \nn
	&=-i\,f_V^\perp  \int_0^1du\,e^{i\,u\,p\cdot x}\,\Bigg\{ \epsilon^*_{[\mu}\,p_{\nu]}  \Big[\phi_2^\perp(u)+\frac{m_V^2\,x^2}{16}\,\phi_4^\perp(u)\Big] \nonumber\\
	&\quad -p_{[\mu}\,x_{\nu]}\,\epsilon^*\cdot x\,m_V^2 \Big[\hat{\hat{\phi}}_3^\parallel(u) -\frac{1}{2} \hat{\hat{\phi}}_2^\perp(u) -\frac{1}{2} \hat{\hat{\psi}}_4^\perp(u) \Big] 
	-\epsilon^*_{[\mu}\,x_{\nu]}\frac{i}{2}\, m_V^2 \big[\hat{\psi}_4^\perp(u) -\hat{\phi}_2^\perp(u)\big] \Bigg\} \nn
	&\equiv-i\,f_V^\perp  \int_0^1du\,e^{i\,u\,p\cdot x}\,\Bigg\{ \epsilon^*_{[\mu}\,p_{\nu]}  \Big[\phi_2^\perp(u)+\frac{m_V^2\,x^2}{16}\,\phi_4^\perp(u)\Big] \nn
	& \hspace{4cm} -p_{[\mu}\,x_{\nu]}\,\epsilon^*\cdot x\,m_V^2\,\phi_3^{\parallel,(2)}  -\epsilon^*_{[\mu}\,x_{\nu]}\frac{i}{2}\, m_V^2 \psi_4^{\perp,(1)}(u)\Bigg\}\,, \label{DAs2}\\
	&\langle V(p,\epsilon)|\bar{q}_1(x)\,\gamma_\mu\gamma_5 \,q_2(0)|0\rangle =\frac{1}{4}\,f_V^\parallel\,m_V\, \epsilon_{\mu\nu\rho\sigma} \epsilon^{*\nu} p^\rho x^\sigma \int_0^1du\,e^{i\,u\,p\cdot x}\,\Big[\psi_3^\perp(u)+ \frac{m_V^2 x^2}{16} \psi_5^\perp(u)\Big]\,, \\
	&\langle V(p,\epsilon) | \bar{q}_1(x) q_2(0) |0\rangle =-\frac{i}{2} f_V^\perp \, \epsilon^*\cdot x\, m_V^2 \int_0^1du \,e^{iup\cdot x} \,\psi_3^\parallel(u) \,. 
\end{align}
In Eq.~(\ref{DAs1}) and (\ref{DAs2}), the apparent kinematic singular factors $1/(p \cdot x)$ have been eliminated by integrating the DAs. This procedure requires the following normalization conditions to be  satisfied at the boundaries,
\begin{align}
	\phi_2^{\parallel,(1)}(1) =\phi_4^{\parallel,(1)}(1) = \psi_4^{\parallel,(1)}(1) =\psi_4^{\parallel,(2)}(1)=\phi_3^{\parallel,(1)}(1) =\phi_3^{\parallel,(2)}(1)= \psi_4^{\perp,(1)}(1)=0 \,, 
\end{align}
and the integrated functions are defined as
\begin{align}
	\hat{\phi}(u)=\int_0^u dv\,\phi(v) \,, \quad \hat{\hat{\phi}}(u)=\int_0^u dv\int_0^v dv'\,\phi(v') \,.
\end{align}

The explicit DAs utilized in this work are given by~\cite{Bharucha:2015bzk,Ball:2007rt}
\begin{align}
	\phi_2^\parallel(u)&=6u\bar{u}\Big(1+ 3\xi^2 \, a_1^\parallel  +\frac{3}{2}(5\xi^2-1) \, a_2^\parallel \Big) \,, \\
	\phi_2^\perp(u)&=6u\bar{u}\Big(1+ 3\xi^2 \, a_1^\perp + \frac{3}{2}(5\xi^2-1)\, a_2^\perp\Big) \,, \\
	\psi_{3}^\parallel(u)  &=  6 u\bar u \Big[ 1 + \Big(
	\frac{1}{3}a_1^\perp + \frac{5}{3} \kappa_{3}^\perp\Big)
	C_1^{3/2}(\xi)  + \Big( \frac{1}{6}a_2^\perp  + \frac{5}{18}
	\omega_{3}^\perp\Big) C_2^{3/2}(\xi)
	-\frac{1}{20}\lambda_{3}^\perp C_3^{3/2}(\xi)\Big] \nn
	&\quad + 3\, \delta_- \Big[ u\bar u \left(9 a_1^\parallel + 10\xi a_2^\parallel\right)
	\bar u \ln \bar u \left(1+3 a_1^\parallel + 6 a_2^\parallel\right)
	- u \ln u \left(1-3 a_1^\parallel + 6 a_2^\parallel\right)\Big] \nn
	&\quad +3\,\delta_+\Big[
	u\bar u \left(1 + 2 \xi a_1^\parallel + 3 (7-5u\bar u) a_2^\parallel\right) 
	+ \bar u \ln \bar u \left(1+3 a_1^\parallel + 6 a_2^\parallel\right)  \nn
	&\quad + u \ln u \left(1-3 a_1^\parallel + 6 a_2^\parallel\right)\Big]\,,  \\
	\psi_{3}^\perp(u)  &=  6 u\bar u \Big[ 1 + \Big(
	\frac{1}{3}a_1^\parallel + \frac{20}{9} \kappa_{3}^\parallel\Big)
	C_1^{3/2}(\xi)  + \Big( \frac{1}{6}a_2^\parallel  + \frac{10}{9}
	\zeta_{3}^\parallel+\frac{5}{12} \omega_3^\parallel-\frac{5}{24} \tilde{\omega}_3^\parallel \Big) C_2^{3/2}(\xi) \nn
	&\quad +\Big(\frac{1}{4} \tilde{\lambda}_3^\parallel -\frac{1}{8}\lambda_{3}^\parallel\Big) C_3^{3/2}(\xi)\Big] 
	+6\,\tilde{\delta}_+\Big[
	u\bar u \left(2 + 3 \xi a_1^\perp + 2 (11-10u\bar u) a_2^\perp\right) \nn 
	&\quad + \bar u \ln \bar u \left(1+3 a_1^\perp + 6 a_2^\perp\right)  
	+ u \ln u \left(1-3 a_1^\perp + 6 a_2^\perp\right)\Big]
	+6\,\tilde{\delta}_- \Big[ u\bar u \left(9 a_1^\perp + 10\xi a_2^\perp\right)\nn
	&\quad + \bar u \ln \bar u \left(1+3 a_1^\perp + 6 a_2^\perp\right)- u \ln u
	\left(1-3 a_1^\perp + 6 a_2^\perp\right)\Big],\\
	\phi_{3}^\perp(u)
	&= \frac{3}{4}\,(1+\xi^2) + \frac{3}{2}\,\xi^3
	a_1^{\|} + \left( \frac{3}{7}\,a_2^{\|} + 5
	\zeta_{3}^{\|}\right) (3\xi^2-1) 
	+ \Big(5\kappa_{3}^{\|}- \frac{15}{16}\,\lambda_{3}^{\|} 
	+\frac{15}{8}\,\widetilde{\lambda}_{3}^{\|} \Big) \xi(5\xi^2-3) \nn
	&\quad + \Big(\frac{9}{112}\,a_2^{\|} +
	\frac{15}{32}\,\omega_{3}^{\|} -
	\frac{15}{64}\,\widetilde\omega_{3}^{\|} \Big)(35\xi^4-30\xi^2+3) 
	+\frac{3}{2}\,\tilde{\delta}_+ \Big[2 + 9 \xi a_1^\perp 
	+ 2 (11-30 u {\bar u}) a_2^\perp \nn
	&\quad +\Big(1-3 a_1^\perp + 6 a_2^\perp\Big) \ln u + \Big(1+3 a_1^\perp + 6 a_2^\perp\Big)\ln {\bar u} \Big] 
	+\frac{3}{2}\,\tilde{\delta}_- \Big[ 2\xi + 9 (1-2u {\bar u}) a_1^\perp \nn
	&\quad + 2 \xi (11-20 u {\bar u}) a_2^\perp 
	+(1+3 a_1^\perp) + 6 a_2^\perp) \ln {\bar u}  
	- (1-3 a_1^\perp + 6 a_2^\perp)	\ln u \Big] \,,\\
	\phi_4^{\parallel}(u)&=24 u^2\bar{u}^2\,, \quad  \psi_4^{\parallel,(2)}(u)=-\frac{3}{2} u^2\bar{u}^2\,, \quad 	\phi_4^\perp(u)=24 u^2\bar{u}^2\,, \quad \psi_4^{\perp,(1)}(u)=0\,, \label{DAs-expand}
\end{align}
with $\xi=2u-1$. And the mass ratios are defined as $\delta_{\pm}=f_{V}^\parallel(m_{q_2}\pm m_{q_1})/(f_{V}^\perp m_{V})$,  $\tilde{\delta}_{\pm}=f_{V}^\perp(m_{q_2}\pm m_{q_1})/(f_{V}^\parallel m_{V})$. The parallel and transverse decay constants, $f_V^\parallel$ and $f_V^\perp$, are parameterized via
\begin{align}
	\langle V(p,\epsilon)|\bar{q}_1\,\gamma_\mu\,q_2|0\rangle= f_V^\parallel \,m_V\,\epsilon^*_\mu\,, \quad  \langle V(p,\epsilon) |\bar{q}_1\,\sigma_{\mu\nu}\,q_2|0\rangle=-i\,f_V^\perp\,(\epsilon^*_\mu\,p_\nu-\epsilon^*_\nu\,p_\mu)\,.
\end{align}

It is instructive to clarify our treatment of the higher-twist DAs in comparison to recent literature~\cite{Wang:2020yvi,Jin:2024zyy,Jiang:2024equ,Aliev:2021cjt}. While incorporating complete non-asymptotic expressions for higher-twist DAs can capture broader phenomenological effects, integrating these terms by parts generically yields non-vanishing boundary (surface) contributions. Discarding these boundary terms leaves uncanceled kinematic singularities, breaking the mathematical consistency of the light-cone expansion. Furthermore, a consistent cancellation of leading-order kinematic singularities strictly requires the inclusion of twist-5 DAs~\cite{Bharucha:2015bzk}. To maintain theoretical rigor and prevent uncontrollable systematic truncation errors, we utilize the complete non-asymptotic expressions exclusively for regular terms that are free of kinematic singularities. For singular terms that require integration by parts, we adhere to the asymptotic forms, ensuring that all boundary terms identically vanish.

For the three-particle distributions, we retain the complete genuine twist-3 DAs to capture the leading multi-parton correlation effects. For the twist-4 contributions, we adopt the Wandzura-Wilczek approximation, formally setting the genuine quark-antiquark-gluon coupling parameters (e.g., $\zeta_4$) to zero. This removes the three-particle twist-4 DAs and reduces the two-particle ones to asymptotic forms, yielding a consistent $\mathcal{O}(\lambda^2)$ truncation. The explicit definitions for the three-particle DAs are given by~\cite{Ball:2007rt}
\begin{align}
	\langle V(p,\epsilon)|\bar{q}_1(x)g_s\,G_{\mu\nu}(vx)\gamma_\alpha\, q_2(0)|0\rangle 
	&= if_V^\parallel\,m_V \int[D\alpha] \,e^{i\,\alpha_v\,p\cdot x}\, p_\alpha\,p_{[\nu}\, \epsilon^{*\perp}_{\mu]} \Phi_3^\parallel(\alpha)\,, \\
	\langle V(p,\epsilon)|\bar{q}_1(x)g_s\tilde{G}_{\mu\nu}(vx) \gamma_\alpha\gamma_5q_2(0)|0\rangle 
	&= -f_V^\parallel m_V\int[D\alpha]\,e^{i\,\alpha_v\,p\cdot x} \, p_\alpha\,p_{[\nu}\, \epsilon^{*\perp}_{\mu]}\, \tilde{\Phi}_3^\parallel(\alpha)   \,, \\
	\langle V(p,\epsilon)|\bar{q}_1(x)g_sG_{\mu\nu}(vx) \sigma_{\alpha\beta}q_2(0)|0\rangle 
	&= f^\perp_Vm_V^2\int[D\alpha] \,e^{i\,\alpha_v\,p\cdot x}\, \frac{\epsilon^*\cdot x}{2\,p\cdot x} 
	\big[ p_\mu p_{[\alpha}\,g^\perp_{\beta]\nu} \nn
	&\hspace{3.5cm}-(\mu\leftrightarrow \nu) \big] \Phi_3^\perp(\alpha) \,,
\end{align}
where the integration measure over the momentum fractions is defined as
\begin{align}
	\int[D\alpha]=\int_0^1 d\alpha_q \int_0^1 d\alpha_{\bar{q}} \int_0^1 d\alpha_g \, \delta(1-\alpha_q-\alpha_{\bar{q}} -\alpha_g)  \,, \quad  \alpha_v= \alpha_q+v\,\alpha_g\,. 
\end{align}
The explicit expressions  utilized in this work read
\begin{align}
	\Phi_3^\parallel(\alpha_i)&=360\alpha_q\alpha_{\bar{q}}\alpha_g^2 \Big[\kappa_3^\parallel +\omega_3^\parallel(\alpha_q-\alpha_{\bar{q}}) +\lambda_3^\parallel \frac{1}{2} (7\alpha_g-3)\Big] \,, \\
	\tilde{\Phi}_3^\parallel(\alpha_i)&=-360\alpha_q\alpha_{\bar{q}}\alpha_g^2 \Big[\zeta_3^\parallel +\tilde{\lambda}_3^\parallel (\alpha_q-\alpha_{\bar{q}}) +\tilde{\omega}_3^\parallel \frac{1}{2} (7\alpha_g-3)\Big] \,.
\end{align}

Finally, the scale dependence of the Gegenbauer moments $a_n^{\parallel(\perp)}$ at NLL accuracy reads
\begin{align}
	&a_n^\parallel(\mu)=E_{V,n}^{\rm NLO}(\mu,\mu_0) a_n^\parallel(\mu_0) +\frac{\alpha_s(\mu)}{4\pi} \sum_{k=0}^{n-2} E_{V,n}^{\rm LO}(\mu,\mu_0) d_{V,n}^k(\mu,\mu_0) a_k^\parallel(\mu_0) \,,\\
	&\big[f_V^\perp(\mu)a_n^\perp(\mu)\big]=E_{T,n}^{\rm NLO}(\mu,\mu_0) \big[f_V^\perp(\mu_0)a_n^\perp(\mu_0)\big] \nn &\hspace{3.5cm}+\frac{\alpha_s(\mu)}{4\pi} \sum_{k=0}^{n-2} E_{T,n}^{\rm LO}(\mu,\mu_0) d_{T,n}^k(\mu,\mu_0) \big[f_V^\perp(\mu)a_k^\perp(\mu_0)\big]\,.
\end{align}
The explicit expressions for the evolution functions $E_{V(T),n}^{\rm (N)LO}$, the mixing coefficients $d_{V(T),n}^k$, and the corresponding anomalous dimensions can be found in Refs.~\cite{Wang:2017ijn,Agaev:2010aq,Belitsky:2005qn}. The renormalization group evolution of the twist-3 parameters at LL accuracy is given by \cite{Ball:2007rt}
\begin{align}
	\omega_3^\perp(\mu)&=R^{(C_A+17C_F/6)/\beta_0}  \omega_3^\perp(\mu_0)\,, \nn \kappa_3^\parallel[\zeta_3^\parallel](\mu) &=R^{(3C_A+C_F/3)/\beta_0} \kappa_3^\parallel[\zeta_3^\parallel](\mu_0) \,, \nn
	\begin{pmatrix}
		\frac{2}{3} \omega_3^\parallel-\tilde{\omega}_3^\parallel \\
		\frac{2}{3} \omega_3^\parallel+\tilde{\omega}_3^\parallel
	\end{pmatrix}
	(\mu)&=R^{\Gamma_3^+/\beta_0} 
	\begin{pmatrix}
		\frac{2}{3} \omega_3^\parallel-\tilde{\omega}_3^\parallel \\
		\frac{2}{3} \omega_3^\parallel+\tilde{\omega}_3^\parallel
	\end{pmatrix}
	(\mu_0)\,, \nn
	\begin{pmatrix}
		\frac{2}{3} \tilde{\lambda}_3^\parallel+\lambda_3^\parallel \\
		\frac{2}{3} \tilde{\lambda}_3^\parallel-\lambda_3^\parallel
	\end{pmatrix}
	(\mu)&=R^{\Gamma_3^-/\beta_0} 
	\begin{pmatrix}
		\frac{2}{3} \tilde{\lambda}_3^\parallel+\lambda_3^\parallel \\
		\frac{2}{3} \tilde{\lambda}_3^\parallel-\lambda_3^\parallel
	\end{pmatrix}
	(\mu_0)\,,
\end{align}
with the ratio $R=\alpha_s(\mu)/\alpha_s(\mu_0)$ and the anomalous dimension
\begin{align}
	\Gamma_3^+=
	\begin{pmatrix}
		\frac{7}{3}C_A+\frac{8}{3} C_F & -\frac{2}{3}C_A+\frac{2}{3} C_F \\
		-\frac{4}{3}C_A+\frac{5}{3} C_F & 4C_A+\frac{1}{6} C_F
	\end{pmatrix}
	\,,\quad  \Gamma_3^-=
	\begin{pmatrix}
		4C_A+\frac{1}{6} C_F & -\frac{4}{3}C_A+\frac{5}{3} C_F \\
		-\frac{2}{3}C_A+\frac{2}{3} C_F & \frac{7}{3}C_A+\frac{8}{3} C_F
	\end{pmatrix}\,.
\end{align}

The numerical values of the fundamental hadronic parameters governing the light vector meson DAs employed in our analysis are summarized in Table~\ref{para_DA}. 
\begin{table}[h]
	\centering 
	\renewcommand{\arraystretch}{1.0}
	\begin{tabular}{|c|cccc|} 
		\hline 
		Parameter & $\rho$ & $\omega$ & $K^*$ & $\phi$ \\
		\hline
		$f^\parallel$ [MeV] & $213(5)$ & $197(8)$ & $204(7)$ & $233(4)$ \\
		$f^\perp$ [MeV] & $160(7)$ & $148(13)$ & $159(6)$ & $191(4)$ \\
		$a_2^\parallel$ & $0.17(7)$ & $0.15(12)$ & $0.16(9)$ & $0.23(8)$ \\
		$a_2^\perp$ & $0.14(6)$ & $0.14(12)$ & $0.10(8)$ & $0.14(7)$ \\
		$a_1^\parallel$ & $-$ & $-$ & $0.06(4)$ & $-$ \\
		$a_1^\perp$ & $-$ & $-$ & $0.04(3)$ & $-$ \\
		$\omega_3^\perp$ & $0.55(25)$ & $0.55(25)$ & $0.3(1)$ & $0.20(8)$ \\
		$\omega_3^\parallel$ & $0.15(5)$ & $0.15(5)$ & $0.10(4)$ & $0.09(3)$ \\
		$\tilde{\omega}_3^\parallel$ & $-0.09(3)$ & $-0.09(3)$ & $-0.07(3)$ & $-0.045(15)$ \\
		$\zeta_3^\parallel$ & $0.030(10)$ & $0.030(10)$ & $0.023(8)$ & $0.024(8)$ \\
		$\lambda_3^\parallel$ & $0$ & $0$ & $-0.008(4)$ & $0$ \\
		$\tilde{\lambda}_3^\parallel$ & $0$ & $0$ & $0.035(15)$ & $0$ \\
		$\kappa_3^\parallel$ & $0$ & $0$ & $0.000(1)$ & $0$ \\
		\hline
	\end{tabular} 
	\caption{The hadronic parameters of vector meson DAs. Scale dependent quantities (except for $f^\parallel$) are evaluated at $\mu_0=1$ GeV. The parameters $f^{\parallel,\perp}$, $a_{1,2}^{\parallel,\perp}$ are obtained in~\cite{Bharucha:2015bzk}, and the twist-3 DA parameters are taken from~\cite{Ball:2007rt}.}
	\label{para_DA}
\end{table}

\section{Explicit expressions for the one-loop Feynman diagrams}
\label{appendix-B}

The analytical results of the one-loop Feynman diagrams in Fig.~\ref{twist-2-NLO} for the partonic QCD matrix elements are presented.

After stripping off the common Lorentz structures and the partonic SCET matrix elements, the loop integrals yield purely scalar correction factors. The exact scalar expressions for the $\mathcal{O}(\alpha_s)$ vertex corrections, quark self-energy, and box diagrams (labeled as $a,b,c,d$ in Fig.~\ref{twist-2-NLO}) for the charge channel read
\begin{align}
	F_g^{(1a)}&=\frac{\alpha_sC_F}{4\pi} \Biggl\{ \Big[\frac{2(1-r_2)}{r_2-r_1} \big(L(r_1)-L(r_2)\big) -1 \Big] \Big(\frac{1}{\epsilon}+L\Big) 
	+\frac{2(1-r_2)}{r_1-r_2} \big[L_2(r_1)-L_2(r_2)  \big]  \nn
	&\quad +\frac{1+r_1^2-2r_2}{r_1(r_1-r_2)} L(r_1) - \frac{(1-r_2)^2}{(r_1-r_2) r_2} L(r_2) -3  \Biggr\} H_g^{(0)}\,, \nn
	F_g^{(1b)}&=F_g^{(1a)}(r_2\to r_3)\,, \nn
	F_g^{(1c)}
	&= - {\alpha_s \, C_F \over 4 \, \pi} \, \frac{1}{1-r_1}\Bigg \{ \Big({1 \over \epsilon} +L\Big)(7 - r_1) -\frac{r_1^2-6r_1+1}{r_1^2}(1-r_1) L(r_1) 
	- \frac{1}{r_1}+3 \Bigg\}\, H_g^{(0)} \,, \nn
	F_g^{(1d)}
	&= -{\alpha_s \, C_F \over 2 \, \pi} \, (1-r_1)\Bigg \{ \bigg[\frac{1-r_1}{(r_1-r_2)(r_1-r_3)}  L(r_1)  -\frac{1-r_2}{(r_1-r_2)(r_2-r_3)}  L(r_2)  \nn 
	&\quad +\frac{1-r_3}{(r_1-r_3)(r_2-r_3)}  L(r_3) \bigg] \Big(\frac{1}{\epsilon} +L \Big) 
	-\frac{1-r_1}{(r_1-r_2)(r_1-r_3)}  \Big[ L_2(r_1)  +\frac{1+r_1}{r_1} L(r_1) \Big] \nn
	&\quad +\frac{1-r_2}{(r_1-r_2)(r_2-r_3)}  \Big[ L_2(r_2)  +\frac{1+r_2}{r_2} L(r_2) \Big] \nn
	&\quad -\frac{1-r_3}{(r_2-r_3)(r_1-r_3)} \Big[ L_2(r_3)  +\frac{1+r_3}{r_3} L(r_3) \Big]  \Bigg\}  H_g^{(0)} \,.
\end{align}
Similarly, the corresponding one-loop matching coefficients for the magnetic channel are given by
\begin{align}
	F_{f}^{(1a)} &=
	\frac{\alpha_sC_F}{4\pi} \Biggl\{ \Big[\frac{2(1-r_2)}{r_2-r_1} \big(L(r_1)-L(r_2)\big) -1 \Big] \Big(\frac{1}{\epsilon}+L\Big) 
	+\frac{2(1-r_2)}{r_1-r_2} \big[L_2(r_1)-L_2(r_2)  \big]  \nn
	&\quad -\frac{1-r_1^2}{r_1(r_1-r_2)} L(r_1) - \frac{1-r_2^2}{(r_1-r_2) r_2} L(r_2) -3  \Biggr\}\,H_f^{(0)} \,,\nn
	F_{f}^{(1b)} &=
	\frac{\alpha_sC_F}{4\pi} \Biggl\{ \Big[\frac{2(1-r_3)}{r_3-r_1} \big(L(r_1)-L(r_3)\big) -1 \Big] \Big(\frac{1}{\epsilon}+L\Big) 
	+\frac{2(1-r_3)}{r_1-r_3} \big[L_2(r_1)-L_2(r_3)  \big]  \nn
	&\quad -\frac{(1-r_1)(r_1+2r_3-1)}{r_1(r_1-r_3)} L(r_1) - \frac{(1-r_3)(1-3r_3)}{(r_1-r_3) r_3} L(r_3) -4  \Biggr\}\, H_f^{(0)} \,,\nn
	F_{f}^{(1c)}&= -\frac{\alpha_s  C_F}{2\pi} \frac{1}{1-r_1}\Big[(r_1+2)\Big(\frac{1}{\epsilon}+L\Big) +\frac{1-r_1^2}{r_1} L(r_1) +2(1+r_1)\Big]\, H_f^{(0)}\,, \nn
	F_{f}^{(1d)}&=0\,.
\end{align}

\end{document}